\documentclass[aps,pra,superscriptaddress,twocolumn]{revtex4-2}
\usepackage{bm}
\usepackage{amsmath}
\usepackage{amsthm}
\usepackage{amsfonts}
\usepackage{amssymb}
\usepackage{latexsym}
\usepackage[utf8]{inputenc}		
\allowdisplaybreaks
\begin{document}
\preprint{Version 1.0}

\title{Quantum electrodynamics of  two-body systems with arbitrary masses}

\author{Jacek Zatorski}
\affiliation{Faculty of Physics, University of Warsaw, Pasteura 5, 02-093 Warsaw, Poland}
\author{Vojt\v{e}ch Patk\'o\v{s}}
\affiliation{Faculty of Mathematics and Physics, Charles University,  Ke Karlovu 3, 121 16 Prague
2, Czech Republic}
\author{Krzysztof Pachucki}
\affiliation{Faculty of Physics, University of Warsaw, Pasteura 5, 02-093 Warsaw, Poland}

\date{\today}

\begin{abstract} 
We perform a calculation of quantum electrodynamics effects in excited states with $l>1$ of arbitrary two-body systems
up to $\alpha^6\,\mu$ order. The obtained results are valid for  hadronic atoms, as long as the strong interaction effects are negligible.
We demonstrate that for circular states with $l\sim n$, the effective expansion parameter is $Z\,\alpha/n$, which extends the applicability of the derived formulas to heavy ions.
Moreover, inclusion of higher-order terms is feasible, which indicates that accurate measurements of excited states of, for example, muonic or antiprotonic hydrogenic ions
can be used to determine the fundamental constants and to search for the existence of yet unknown long-range interactions.  
\end{abstract}

\maketitle

\section{Introduction \label{section:introduction}}

Finite nuclear mass corrections to atomic energy levels are often the main limiting factor of theoretical predictions, 
for example in the isotope shift of regular atoms, or transition frequencies in exotic atoms such as  muonium $e^-\,\mu^+$ \cite{crivelli_22}. 
It is because there is no fundamental equation for a two-body system with arbitrary masses which accounts for relativity.
This is in contrast with the Dirac equation, which holds when one of the particles is infinitely heavy. Although there is an exact formula \cite{GrotchYennie}, which gives
the leading $m/M$ correction valid for an arbitrary nuclear charge, the more general result for an arbitrary mass ratio is not known. 
In the alternative approach one employs QED and expansion in the fine structure constant $\alpha$ to derive an analytic formula 
for nuclear recoil corrections in consequent orders in $\alpha$.
Here we demonstrate that one can derive exact formulas for $\alpha$ expansion coefficients for excited rotational $l>1$ states, 
which are valid for arbitrary masses of its constituents.
We perform calculations up to the order $\alpha^6\,\mu$ and indicate the possibility of further improvements. This would allow for extremely accurate
 theoretical predictions for excited hydrogenic states as long as the contact interactions, e.g. from strong forces, are negligible.
 A good example are excited states of antiprotonic atoms, like $\bar p\,p$ or $\bar p\,\alpha$.  If corresponding measurements are available,
 this would allow for more accurate determination of fundamental constants or tests of the existence of unknown long-range interactions,
 similarly to precision tests performed with antiprotonic helium \cite{Widmann1, HayanoHori, Hori2005}, but much more accurate.
 
 Several accurate results have been obtained for excited states of hydrogenic systems  including two-loop QED in the infinite nuclear mass limit, and leading $m/M$ corrections.
 This, in combination with $\mu$H Lamb shift measurement, allowed for the determination of the Rydberg constant and the proton charge radius. 
 The measurement of magnetic interactions in hydrogen-like carbon is currently the best determination of the electron mass in atomic mass units.
 The measurement of muonium $1S-2S$ will in the future provide the $\mu$ mass. If accurate measurements are available for hydrogenic rotational states \cite{NIST}, 
 this would allow not only for more accurate determination of fundamental constants, but also will allow for the search of long-range interactions.
 For all these examples, accurate theoretical predictions of energy levels are necessary, and in this work we demonstrate the availability 
 of highly accurate results for rotational levels. 
 
The rest of this paper is organized as follows. In Section II we outline the method of calculation. Section III contains details of the calculation of the leading relativistic correction 
to energy. In Section IV we present the leading QED correction to energy levels, and Section V is devoted to higher-order QED corrections to the energy levels,
followed by a summary in Section VI.

\section{Expansion of energy in powers of the fine structure constant $\alpha$}
In this work we consider two-body systems consisting of a negatively charged particle of mass $m_1$ and a positively charged nucleus of mass $m_2$.
From the theoretical perspective the relevant atomic transitions are those between states of the angular momentum number $l \approx n$. 
For that reason, during calculations we will omit contributions coming from operators containing short-range interactions, namely those of the nuclear size.
We will neglect also  (electronic) vacuum polarization, which, however, can be included separately, and its significance depends on $n$, $l$, and constituent masses. 

As a consequence, the energy of a two-body system with masses $m_1, m_2$, charges $e_1, e_2$, spins $s_1, s_2$,
and g-factors $g_1, g_2$ can be expressed as a series in powers of the fine structure constant $\alpha$,
\begin{align}
E(\alpha) =  E^{(0)} + E^{(2)} +  E^{(4)} + E^{(5)} +  E^{(6)} + o(\alpha^7)\,,
\end{align}  
where each individual term $E^{(j)}$ is of the order $\alpha^j$. In particular,
\begin{align}
E^{(0)} = m_1+m_2\,,
\end{align}
and $E^{(2)}$ is the eigenvalue of the nonrelativistic two-body Hamiltonian $H_0 = H^{(2)}$ in the center of mass frame,
\begin{align}
H_0 = \frac{p^2}{2\,\mu} + \frac{e_1\,e_2}{4\,\pi}\,\frac{1}{r}\,.
\end{align}
When we set $e_1 = -e, e_2 = Z\,e$, it is equal to
\begin{equation}
E^{(2)} = E_0 = -\frac{(Z\,\alpha)^2\,\mu}{2\,n^2}\,,
\end{equation}
with $\mu = m_1\,m_2/(m_1+m_2)$ being the reduced mass.

\section{Leading relativistic correction $E^{(4)}$ }
$E^{(4)}$ is the leading relativistic correction, namely the expectation value of the Breit Hamiltonian $H^{(4)}$ with the nonrelativistic wave function,
\begin{widetext}
\begin{align}
H^{(4)} =&\ -\frac{\vec p^{\,4}}{8\,m_1^3}\  -\frac{\vec p^{\,4}}{8\,m_2^3}
+ \frac{e_1\,e_2}{4\,\pi}\,\biggl\{
\frac{1}{2\,m_1\,m_2}\,p^i\,
\biggl(\frac{\delta^{ij}}{r}+\frac{r^i\,r^j}{r^3}\biggr)\,p^j
+\frac{g_1\,g_2}{4\,m_1\,m_2}\,\biggl[\frac{s_1^i\,s_2^j}
{r^3}\,\biggl(\delta^{ij}-3\,\frac{r^i\,r^j}{r^2}\biggr)
-\frac{8\,\pi}{3}\,\vec s_1\vec s_2\,\delta^{(3)}(\vec r)\biggr]
\nonumber \\ &\
-\frac{\vec r\times\vec p}{2\,r^3} \cdot\biggl[
\frac{g_1}{m_1\,m_2}\,\vec s_1 
+\frac{g_2}{m_1\,m_2}\,\vec s_2
+\frac{(g_2-1)}{m_2^2}\,\vec s_2 
+\frac{(g_1-1)}{m_1^2}\,\vec s_1\biggr]\biggr\}
 - \frac{e_{1} e_{2}}{6} \bigl( \left\langle r_{E}^2\right\rangle_{1} + \left\langle r_{E}^2\right\rangle_{2} \bigr) \, \delta^{(3)}(\vec{r})\,, \label{Breitgeneral}
\end{align}
\end{widetext}
where $\vec r = \vec r_{12}$, $\vec p = \vec p_1 = -\vec p_2$, 
and we neglect the quadrupole moment of any of particles. 
The g-factor is related to the magnetic moment anomaly by  $g = 2(1+\kappa)$. 
For the proton it amounts to $g_p = 5.585\,694\,689\,3(16)$, and for the antiproton it is the same.
For the case of a spin $1/2$ point-like particle the charge radius is
\begin{align}
\left\langle r_{E}^2\right\rangle =&\ \frac{3}{4\,m^2}\,.
\end{align}
For a spinless particle it is $\left\langle r_{E}^2\right\rangle = g = 0$.
Let us note that Hamiltonian (\ref{Breitgeneral}) does not account for any potential annihilation effects, which would be present, for example, in positronium.
It also does not account for strong interaction effects, which are present for hadronic particles.

From the Hamiltonian in Eq.~(\ref{Breitgeneral}) one obtains $E^{(4)}$ by taking the expectation value with the nonrelativistic eigenfunction of $H_0$.
The result for a state with the principal quantum number $n$ and the angular momentum $l$ is
\begin{widetext}
\begin{align}
E^{(4)} =& \mu^3 (Z\alpha)^4\bigg\{\frac{1}{8\, n^4}\,\biggl(\frac{3}{\mu^2}-\frac{1}{m_1\,m_2}\biggr)
-\frac{1}{\mu^2(2l+1) \, n^3}
+\frac{2\,\delta_{l0}}{3\,n^3} \bigl( \left\langle r_{E}^2\right\rangle_{1} + \left\langle r_{E}^2\right\rangle_{2} \bigr)
+\frac{\delta_{l0}}{m_1\, m_2\,n^3}\nonumber\\
&+\frac{2}{l(l+1) (2l+1) n^3}\bigg[\vec L\cdot\vec s_1\bigg(\frac{1+2\kappa_1}{2m_1^2}+\frac{1+\kappa_1}{m_1 m_2}\bigg)
+ \vec L\cdot\vec s_2\bigg(\frac{1+2\kappa_2}{2m_2^2}+\frac{1+\kappa_2}{m_1 m_2}\bigg)\nonumber\\
& - \frac{6(1+\kappa_1)(1+\kappa_2)}{m_1 m_2\,(2\,l - 1) (2\,l+3)}s_1^i s_2^j (L^i L^j)^{(2)}\bigg]
+\frac{8\,\delta_{l0}}{3m_1 m_2\,n^3}(1+\kappa_1)(1+\kappa_2)\,\vec s_1\cdot\vec s_2 
\bigg\}\,.\label{E4general}
\end{align}
\end{widetext}
In the last equation we have introduced a symmetric traceless tensor $(L^i L^j)^{(2)}$, which is defined as
\begin{equation}
(L^i L^j)^{(2)} = \frac12\big( L^i L^j + L^j L^i) - \frac{\delta^{ij}}{3}\vec L^2\,.
\end{equation}
Expression (\ref{E4general}) is valid for arbitrary spin of both particles, point-like or hadronic.
We will now explore particular cases of the general result in Eq. (\ref{E4general}).

\subsection{Dirac limit of an infinitely heavy and point-like nucleus}

For an infinitely heavy $M=m_2$ and point-like nucleus we set
$m=m_1$, $\vec s = \vec s_1$, and $\kappa = \kappa_1$ to obtain
\begin{align}
&E^{(4,0)} = m\,(Z\alpha)^4\bigg\{\frac{3}{8\, n^4}
-\frac{1}{(2\,l+1) \, n^3}
\nonumber\\&
+\frac{1+2\kappa}{l\,(l+1)\,(2\,l+1)\, n^3}\,\vec L\cdot\vec s
+\frac{2\,\delta_{l0}}{3\,n^3}\,m^2 \left\langle r_{E}^2\right\rangle\bigg\}\,.
\end{align}
If the first particle is also point-like, then
\begin{align}
E^{(4,0)} =& m\,(Z\alpha)^4\bigg\{\frac{3}{8\, n^4}
-\frac{1}{(2\,l+1) \, n^3}
\nonumber\\&
+\frac{1}{l\,(l+1)\,(2\,l+1) n^3}\,\vec L\cdot\vec s
+\frac{\delta_{l0}}{2\,n^3}\bigg\}\,.
\end{align}
Using the total angular momentum $\vec J = \vec L+\vec s$,  it can be rewritten in the form
\begin{equation}\label{13}
E^{(4,0)} = m\,(Z\alpha)^4\bigg\{\frac{3}{8\, n^4}
-\frac{1}{(2j+1) \, n^3}\bigg\}\,.
\end{equation}
This can be compared to  the result obtained from the Dirac equation, which is
\begin{align}\label{DiracRes}
E_D = &\ m\, f(n,j)\,,\\
f(n,j) =&\  \Bigg[1+\frac{(Z\alpha)^2}{\big(n-\big(j+\frac12\big)+\sqrt{(j+\frac12)^2-(Z\alpha)^2}\big)^2}\Bigg]^{-\frac12}\,.
\end{align}
Expanding this up to the order $(Z\alpha)^4$  we get
\begin{equation}
E_D = m\bigg[1-\frac{(Z\alpha)^2}{2n^2} + (Z\alpha)^4\bigg(\frac{3}{8\,n^4} - \frac{1}{(2j+1)\,n^3}\bigg)\bigg]\,.
\end{equation}
The second term is nonrelativistic energy, and the third term is the relativistic correction, in agreement with Eq. (\ref{13}).

\subsection{Atom with a spinless, point nucleus and point, spin-$1/2$ particle}

For the case of a spinless point nucleus of finite mass $M=m_2$ and spin $1/2$ point particle of mass $m=m_1$ with $\vec s=\vec s_1$ we get 
\begin{align}
&E^{(4)} =\frac{ \mu \,(Z\alpha)^4}{n^3}\bigg\{\frac{1}{8\, n}\,\biggl(3-\frac{\mu^2}{m\,M}\biggr) 
-\frac{1}{(2l+1)}
\nonumber\\&
+\bigg(\frac{\mu^2}{m^2}+\frac{2\,\mu^2}{m M}\bigg)\bigg[\frac{\delta_{l0}}{2}
+\frac{1}{l(l+1) (2\,l+1)}\, \vec L\cdot\vec s\bigg]
\bigg\}\,. \label{scalar}
\end{align}
If the spinless nucleus is much heavier than the 
point spin-$1/2$ particle, then the leading recoil correction from Eq.~(\ref{scalar}) equals
\begin{equation}\label{17}
E^{(4,1)} = \frac{m^2}{M}\,(Z\alpha)^4\bigg\{-\frac{1}{2\,n^4}+\frac{1}{(2j+1)\,n^3}\bigg\}\,.
\end{equation}
This is in agreement with the known recoil correction to the Dirac energy \cite{GrotchYennie},
\begin{align}\label{diracrec}
E_{\rm rec}(n,j) = \frac{m}{2\,M}\,\bigl(1-f(n,j)^2\bigr)
\end{align}
Expanding this to the fourth order in $Z\alpha$ we obtain a result in agreement with Eq. (\ref{17}). 

\subsection{Two point spin-1/2 particles with equal mass}

For two point particles with equal mass $m_1=m_2=m$ and spin 1/2 we can define the total spin $\vec S = \vec s_1 + \vec s_2$
and get
\begin{align}
&\ E^{(4)} =  \frac{m}{8} \frac{\alpha^4}{n^3}\,\bigg\{\frac{11}{8\, n}
-\frac{4}{(2l+1)} 
+\frac43\delta_{l0}\,\vec S^2
\nonumber\\&\
+\frac{3}{l(l+1) (2\,l+1)} \bigg[ \vec L\cdot\vec S
- \frac{2\,S^i S^j (L^i L^j)^{(2)}}{(2\,l - 1) (2\,l+3)}\bigg]\bigg\}
\end{align}
where
\begin{align}
\vec L\cdot\vec S =&\ \frac12\big[j(j+1)-l(l+1)-s(s+1)\big]\,,\\
S^i S^j (L^i L^j)^{(2)} =&\ \big(\vec L\cdot\vec S)^2 + \frac12 \vec L\cdot\vec S- \frac13 l(l+1) s(s+1)\,.
\end{align}
This relativistic energy does not account for annihilation terms.

\subsection{Both particles spinless and point-like}

Finally, the case of both particles being spinless and point-like is
\begin{align}
E^{(4)} = \frac{\mu (Z\alpha)^4}{n^3}\bigg[\frac{1}{8\,n}\biggl(3-\frac{\mu^2}{m_1\,m_2}\biggr) -\frac{1}{(2l+1)}
+\delta_{l0}\frac{\mu^2}{m_1\,m_2}\bigg]\,.
\end{align}
In the infinite nuclear mass $M=m_2$ limit
\begin{align}
E^{(4,0)} =&\ m (Z\alpha)^4\bigg\{\frac{3}{8\,n^4} - \frac{1}{(2l+1)\,n^3}\bigg\}\,. 
\end{align}
It has the same form as result (\ref{13}) from the Dirac equation with $j$ replaced by $l$, while the leading recoil correction is
\begin{align}
E^{(4,1)} =&\ \frac{m^2}{M}(Z\alpha)^4\bigg\{-\frac{1}{2n^4} + \frac{1+\delta_{l0}}{(2l+1)\,n^3}\bigg\}\,,
\end{align}
which is different from the case of the Dirac particle in Eq.~(\ref{17}).

\section{Leading QED correction $E^{(5)}$}

QED effects for $l>0$ are partially accounted for in the $g$-factor, which is present in  the Breit-Pauli Hamiltonian $H^{(4)}$.
Additional QED corrections are represented as $E^{(5)}$, and for $l>0$ they have the form 
\begin{align}
E^{(5)} =&\  
-\frac{14\,(Z\,\alpha)^2}{3\,m_1\,m_2}\,\left\langle
\frac{1}{4\,\pi}\,\frac{1}{r^3}\right\rangle-
\frac{2\,\alpha}{3\,\pi}\,\biggl(\frac{1}{m_1}+\frac{Z}{m_2}\biggr)^2
\nonumber \\ &\ \times
\left\langle\,\vec p\,(H_0-E_0)\,
\ln\bigg[\frac{2\,(H_0-E_0)}{\mu(Z\,\alpha)^2}\bigg]\,\vec{p}\right\rangle\,.
\end{align}
The second term is related to the so-called Bethe logarithm $\ln[k_{0}(n, l)]$ by
\begin{align}
\label{def:BetheLogarithm}
	&\ln[k_{0}(n, l)] \equiv \frac{n^3}{2 \mu^3 (Z \alpha)^4}
	\nonumber\\& 
	\times\left\langle \phi\left| \vec{p} \, (H_{0}-E_{0}) \ln\biggl[\frac{2(H_{0}-E_{0})}{\mu (Z\alpha)^2} \biggr] \, \vec{p} \right|\phi \right\rangle 
\end{align}
and has been tabulated for many hydrogenic states.
Let us also write $E^{(5)}$ as a function of quantum numbers $n$ and $l$. Making use of definition~(\ref{def:BetheLogarithm}) and Eq.~(\ref{wzor:r3}) we obtain
\begin{align}
\label{wzor:energyalfa5}
	E^{(5)} =&\ - \frac{7}{3 \pi} \frac{(Z\alpha)^5 \mu^3}{m_{1} m_{2}} \, \frac{1}{l(l+1)(2l+1)\,n^3} \, 
	\nonumber\\ &\
	-\frac{4}{3 \pi} \biggl( \frac{1}{m_1}+\frac{Z}{m_2} \biggr)^2 \frac{\alpha (Z\alpha)^4 \mu^3}{n^3}\,\ln k_{0}(n, l)
\end{align}
in agreement with Eq.~(19) from~Ref. \cite{PachuckiVeita}.

\section{Higher-order QED correction $E^{(6)}$ }

The total contribution of the order $m\,\alpha^6$ to energy can be split into two parts: 
i) the expectation value of the first-order operators with nonrelativistic wave function, and ii) the second-order contribution
induced by the Breit Hamiltonian $H^{(4)}$,
\begin{equation}\label{E6genera}
 E^{(6)} = \langle H^{(6)}\rangle + \langle H^{(4)}\,\frac{1}{(E_0-H_0)'}\,H^{(4)}\rangle\,. 
\end{equation}
Here the prime in $(E_0-H_0)'$ means that we exclude the reference state from the resolvent. The effective operator
$H^{(6)}$ can be derived within the framework of NRQED. 
In general, both the first-order and the second-order contributions separately contain singular operators that cancel
each other when combined together. However, because we are interested in states with $l>1$ then all singularities vanish.
Omitting contact interactions, the NRQED Hamiltonian for a spin $s=0,1/2$ and  finite size particle, from Ref. ~\cite{ZatorskiPachucki2010} is
\begin{align}
	&\ H_{\rm NRQED} = e\,A_{0} + \frac{1}{2\,m} \bigl[\vec \pi^{\,2} - 2\,e\,(1+\kappa)\,\vec{s}\cdot\vec{B} \bigr] 
	\nonumber\\&\
	- \frac{1}{8 m^3} \bigl[\vec\pi^{\,4} - 2e\, \{p^2 \,,\,\vec{s}\cdot\vec{B}\} 
	- 2e\kappa \{\vec{\pi}\cdot\vec{B}, \vec{\pi}\cdot\vec{s} \} \bigr] 
	\nonumber\\ &\
	+ \frac{p^6}{16 m^5} 	
  - \frac{1 + 2\kappa}{4 m^2} \, e \, \vec{s} \cdot (\vec{E}\times\vec{\pi} - \vec{\pi} \times \vec{E} )  
\nonumber\\ &\
- \Bigl(1+\frac{4\,\kappa}{3}\Bigr) \; \frac{3\,e}{16\,m^4} \, \vec s \, \{p^2, \vec{\nabla}A_{0} \times\vec{p} \}
\nonumber\\ &\
+ \frac{\,e}{32\,m^4}\bigg(1+\frac{s_a (s_a+1)}{3}\bigg) [p^2, [p^{2}, A_{0}] ]\nonumber\\ 
&\ - \frac{1}{2}\,\Bigl(\alpha_{E} -\frac{s_a(s_a+1)}{3\,m^3}\Bigr) \, e^2 \, \vec{E}^2 \, , \label{ham:H2}
\end{align}
\vspace{1ex}

\noindent where $[X\,,Y] \equiv X\,Y - Y\,X$ is the commutator of two operators, $\{X\,,\,Y\}\equiv X\,Y+Y\,X$ is the anticommutator, $\vec \pi= \vec p-e\,\vec A$,  
$e$ is the charge, $\kappa$ is the magnetic moment anomaly, which is related to the $g$-factor by $g = 2\,(1+\kappa)$,  
and $e^2\,\alpha_E$ is the static electric dipole polarizability. For point-like particles $\alpha_E = 0$.

Using the NRQED Hamiltonian in Eq. (\ref{ham:H2}) we can derive the effective
operator $H^{(6)}$ for the case of two spinless particles, for one spinless and one spin-1/2 particle, and for two spin-1/2
particles. The derivation of effective operators following Ref. \cite{nrqed}  is presented in Appendix \ref{app:H6}. The general form of $H^{(6)}$ is
\begin{equation}
	H^{(6)} = \sum_{i=0...9} \delta H_{i} \, . \label{hamiltonian:generalH6}
\end{equation}
The individual effective operators are the following
\begin{align}
	\delta H_{0} =& \frac{p^6}{16 m_{1}^5} + \frac{p^6}{16 m_{2}^5} \, ,\label{dH0} \\	
	\delta H_{1} =& \sum_{a} - \frac{3Z\alpha}{16 m_{a}^4}  \biggl(1+\frac{4\kappa_a}{3}\biggr)  \vec L\cdot\vec{s}_{a}\,	
	\{ p^{2}, \frac{1}{r^3} \} \nonumber\\
	&+ \frac{1}{32 m_{a}^{4}}\bigg(1+\frac{s_a(s_a+1)}{3}\bigg) \,[p^2, [p^{2}, V ]] \, ,  \\
\delta H_{2} =& \frac{Z \alpha }{4m_{1}^{2} m_{2}^{2}} (1 + 2\kappa_1)(1 + 2\kappa_2)\nonumber\\
&\, ( \vec{s}_2 \times \vec{p})^i \left( \frac{\delta^{ij}}{r^3} - 3 \frac{r^{i} r^{j}}{r^5} \right) ( \vec{s}_1 \times \vec{p})^j \, , \\
	\delta H_3 =&   \sum_{a} \frac{1}{4m_{a}^{3}} \bigg( \{ p^{2} ,\, \vec{p}_a \cdot e_{a}\vec{\cal{A}}_a\} 
	+ \{p^{2} ,\, \vec{s}_a \cdot \vec{\nabla}_a \times e_{a} \vec{\cal{A}}_a\} \bigg), \\
	\delta H_{4} =& \sum_{a} \frac{e_{a}^2}{2 m_{a}} \, \vec{\cal{A}}_{a}^{2}  \, ,  \\
	\delta H_5 =& \sum_{a} \, \frac{e_{a}^2\,(1+2\kappa_a)}{2 m_{a}^2} \, \vec{s}_{a} \cdot \vec{\cal{E}}_a \times \vec{\cal{A}}_a  \, , \\
		\delta H_{6} =& \,\delta H_{6}^{A} + \delta H_{6}^{B} + \delta H_{6}^{C} \, ,\label{dH6} 
\end{align}
where
\begin{widetext}
\begin{align}
	\nonumber	\delta H_{6}^{A} =& \frac{Z\alpha}{16 m_{1}m_{2}} \bigg\{ \frac{2 Z^2 \alpha^2}{r^3} +
	\frac{iZ\alpha r^{i}}{r^3} \bigg[ \frac{p^{2}}{2m_{2}}, \frac{r^i r^j - 3 \delta^{ij}\,r^2}{r} \bigg] p^j \\
 &- \, p^i \bigg[ \frac{r^i r^j - 3 \delta^{ij}\,r^2}{r}, \frac{p^2}{2m_{1}} \bigg] \frac{iZ\alpha r^{j}}{r^3} - p^{i} \bigg[ \frac{p^2}{2m_{2}}, \bigg[ \frac{r^i r^j - 3 \delta^{ij}\,r^2}{r}, \frac{p^2}{2m_{1}} \bigg] \bigg] p^j 
	\bigg\} + (1 \leftrightarrow 2)	\, ,  \\	 
	\delta H_{6}^{B} =& \frac{Z \alpha}{4 m_{1}m_{2}} \bigg\{ (1+\kappa_{1})\bigg[ \bigg(\vec{s}_{1}\times \frac{\vec{r}}{r}\bigg)^{i}, 
	\, \frac{p^2}{2m_1} \bigg] \frac{iZ\alpha r^{i}}{r^3} 
	- (1+\kappa_{2})\frac{iZ\alpha r^{i}}{r^3} \bigg[ \frac{p^{2}}{2m_{2}}, \bigg(\vec{s}_{2}\times \frac{\vec{r}}{r}\bigg)^{i}\bigg]\nonumber\\&
	 + \,	(1+\kappa_{1}) \bigg[ \frac{p^2}{2 m_2}, \bigg[ \bigg(\vec{s}_{1}\times \frac{\vec{r}}{r}\bigg)^{i}, \, \frac{p^2}{2m_1} \bigg] \bigg] p^i 
	 + (1+\kappa_{2}) p^i \bigg[ \frac{p^2}{2m_2}, \bigg[ \bigg(\vec{s}_{2}\times \frac{\vec{r}}{r}\bigg)^{i}, \, \frac{p^2}{2m_1} \bigg] \bigg] \bigg\} 
	  + (1 \leftrightarrow 2) \, , \\
	\delta H_{6}^{C} =& - \frac{Z \alpha(1+\kappa_{1})(1+\kappa_{2})}{4 m_{1}^{2}m_{2}^{2}} \bigg[ p^2, \bigg[p^2, \, \vec{s}_{1}\vec{s}_{2}\frac{2}{3r} +  s_{1}^{i}s_{2}^{j}\frac{1}{2r}\bigg( \frac{r^i r^j}{r^2} - \frac{\delta^{ij}}{3} \bigg) \bigg] \bigg] 	\, ,  
	\end{align}
	\end{widetext}
	\begin{align}
	&\delta H_7 = \sum_{a} \frac{e_{a}^2\,(1+2\kappa_a)}{2 m_{a}^2} \; \vec{s}_a \cdot \vec{\cal{E}}_a \times \vec{\cal{A}}_a  \nonumber\\&
	+ \frac{ie_{a}\,(1+2\kappa_a)}{8 m_{a}^3} \, \big[ \, \vec{\cal{A}}_a \cdot(\vec{p}_{a} \times \vec{s}_{a}) 
	+ (\vec{p}_{a} \times \vec{s}_{a})\cdot \vec{\cal{A}}_a, p_{a}^2 \, \big] \, ,  \nonumber\\&\\
	&\delta H_{8} = - \frac{1}{2} \sum_{a} \bigg(\alpha_{Ea}-\frac{s_a(s_a+1)}{3\,m_a^3}\bigg) \, \frac{Z^2 \alpha^2}{r^4}	\, ,  \\
	&\delta H_{9} = \sum_{a}
\frac{e_{a}\kappa_{a}}{4 m_{a}^3} \{\vec{p}_{a} \cdot \vec{\nabla}_{a}\times\vec{\cal{A}}_{a}, \vec{p}_a\cdot \vec{s}_a \}	\, . \label{zestawoperatorowgeneralH6}
\end{align}
The static fields ${\cal A}_i$ and ${\cal E}_i$ are defined as
\begin{align}\label{vecA1}
	e_1 {\cal{A}}^{i}_{1} =& - \frac{Z\alpha}{2r} \left( \delta^{ij} + \frac{r^{i} r^{j}}{r^2} \right) \frac{p_{2}^j}{m_2} -
	\frac{Z\alpha}{m_2} \frac{ \left(\vec{s}_{2} \times \vec{r}\right)^i}{r^3} (1+\kappa_2)\;, 	\\
e_2 {\cal{A}}^{i}_{2} =& - \frac{Z\alpha}{2r} \left( \delta^{ij} + \frac{r^{i} r^{j}}{r^2} \right) \frac{p_{1}^j}{m_1} + \frac{Z\alpha}{m_1} \frac{ \left(\vec{s}_{1} \times \vec{r}\right)^i}{r^3}(1+\kappa_1) \; ,  \label{vecA2}\\
	e_{1}\vec{\cal{E}}_1 =& - Z\alpha \frac{\vec{r}}{r^3},\,\,
	e_{2}\vec{\cal{E}}_2 =  Z\alpha \frac{\vec{r}}{r^3} \, .\label{eE}
\end{align} 

In the case when one of the particles is spinless and the other one has spin $1/2$,
the effective operators (\ref{dH0}-\ref{zestawoperatorowgeneralH6}) are changed in such a way that $\vec s_1=s_1=0$ in $\delta H_i$ and also in $e_2\vec{\cal{A}}_2$.
While,  in the case of two spinless particles,
the effective operators (\ref{dH0}-\ref{zestawoperatorowgeneralH6}) are modified by setting $\vec s_1=\vec s_2=s_1=s_2=0$
and the same for static fields $e_a\vec{\cal{A}}_a$. 

The evaluation of the expectation value of $H^{(6)}$ with nonrelativistic wave function
and the second-order contribution with the Breit Hamiltonian for individual cases of particles is then presented in Appendices \ref{app:spin0},
\ref{app:spin12}, and \ref{app:spin1}. The results for energies are presented in the following subsections.

\subsection{Two particles with spins $s_1=s_2=0$}

We will begin with the case when both the positively charged nucleus and negatively charged particle are spinless.
The final result for the energy $E^{(6)}$ is the sum of the first-order and second-order contributions evaluated in Appendix \ref{app:spin0}. 
Defining
\begin{align}
f^{(6)}(n,j) =&  -\frac{5}{16 n^6}+\frac{3}{2 (2 j+1) n^5}-\frac{3}{2 (2 j+1)^2 n^4}
	\nonumber\\&-\frac{1}{(2 j+1)^3 n^3}
\end{align}
we obtain the
correction of the order $m\,\alpha^6$
\begin{widetext}
\begin{align}
   E^{(6)} =& \mu (Z \alpha)^6 \biggl[ 
  f^{(6)}(n,l) + \frac{\mu^2}{m_1\, m_2}\bigg(\frac{3}{16\,n^6}- \frac{8\,l(l+1)-3}{2\, (2l-1)(2l+1)(2l+3)\,n^5}
   + \frac{6}{(2l-1)(2l+1)(2l+3)\, n^3}\bigg)\nonumber\\
   &- \frac{\mu^4}{16\, m_1^2\,m_2^2\,n^6}
   +\frac{2\,\mu^3\,(\alpha_{E1}+\alpha_{E2})}{(2l-1)(2l+1)(2l+3)}\bigg(\frac{1}{n^5}-\frac{3}{l(l+1)\,n^3}\bigg)
  \biggr]	\, . \label{E6spin00}
\end{align}
\end{widetext}
Similarly to $E^{(4)}$, we can explore particular limiting cases of the general result in Eq. (\ref{E6spin00}).
Firstly we perform the limit  where mass $M=m_2$ of the nucleus is infinitely heavy, to obtain the result in agreement with the Klein-Gordon equation:
\begin{equation}
	E^{(6,0)} = m (Z\alpha)^6 \,f^{(6)}(n,l) \, .
\end{equation}
We may also expand Eq.~(\ref{E6spin00}) up to the order $m^2/M$ to obtain the first-order recoil correction
\begin{align}
E^{(6, 1)} =&\   (Z\alpha)^6 \, \frac{m^2}{M} \,f^{(6,1)}_0(n,l)\, ,
\end{align}
where we introduce the function
\begin{align}\label{f610}
	&  f^{(6, 1)}_0(n,l) =   \frac{1}{2 n^6} +\frac{6 - 10 l (l+1)}{(2 l - 1) ( 2 l + 1) (2 l + 3) n^5} \nonumber\\
	&+ \frac{3}{2 (2 l + 1)^2 n^4} 
	+ \frac{3 + 28 l (l+1)}{(2 l - 1) (2 l + 1)^3 (2 l + 3) n^3}\,.	
\end{align}
This concludes the evaluation of the correction to energy for a purely spinless two-body system.

\subsection{Two particles with spins $s_1=0,\,s_2=1/2$}

For the case when one of the particles is spinless while the other one has spin 1/2,
the derivation of the first-order and second-order contributions is presented in Appendix \ref{app:spin12}.
In the resulting expression for energy $E^{(6)}$ we can separate the part
which is already contained in the final result for two spinless particles, Eq.~(\ref{E6spin00}). If we denote the spinless energy (\ref{E6spin00}) as $E^{(6)}_{s_1=s_2=0}$, then 
\begin{align}\label{results0s12}
&E^{(6)} = E^{(6)}_{s_1=s_2=0}  \nonumber\\ &
+ \frac{\mu(Z\alpha)^6}{l(l+1)(2l-1)(2l+1)(2l+3)} \big[ A + B \,\vec L\cdot\vec s_2\big] \,.
\end{align}
With the help of substitutions
\begin{align}
\lambda_0 =&\ \frac{(2\,l-1)(2\,l+3)}{2\,l+1} = 2\,l+1 -\frac{4}{2\,l+1}\,,\\
\lambda_1 =&\ \frac{3}{2l(l+1)} + \frac{4}{(2l+1)^2}\,,\\
\lambda_2 =&\ -\frac{3}{4l^2}+\frac{13}{4l(l+1)}-\frac{3}{4(l+1)^2}+\frac{8}{(2l+1)^2}\,,
\end{align}
and the definition 
\begin{equation}\label{X}
X = \frac{X_5}{n^5} + \frac{X_4}{n^4} + \frac{X_3}{n^3}
\end{equation}
for $X=A,B$
the coefficients $A$ and $B$ in Eq.~(\ref{results0s12}) can be written as 
\begin{align}
A_5 =&\ \frac{l(l+1)}{2}\bigg[g_2^2\frac{\mu^2}{4m_2^2} + (-2-3g_2+g_2^2)\frac{\mu^3}{2m_2^3}+\frac{\mu^4}{m_2^4}\bigg]\,,\nonumber\\
A_4 =&\ \frac{3\lambda_0}{4}\bigg[-g_2^2 \frac{\mu^2}{2m_2^2} + g_2 \frac{\mu^3}{m_2^3} - \frac{\mu^4}{2m_2^4}\bigg]\,,\nonumber\\
A_3 =&\ g_2^2 \bigg(\lambda_1-\frac92\bigg)\frac{\mu^2}{4m_2^2}  + \big(3(2+5g_2-g_2^2) - 2g_2 \lambda_1\big)\frac{\mu^3}{4m_2^3} \nonumber\\&\
+(\lambda_1-9)\frac{\mu^4}{4m_2^4}\, 
\end{align}
and
\begin{widetext}
\begin{align}
B_5 =&\ (2l+1)\lambda_0\bigg[-2 g_2\frac{\mu}{m_2} - 3(g_2+1)\frac{\mu^3}{2m_2^2} + \frac{3\mu^4}{2m_2^4}\bigg]
+ \big(8(5+3g_2) l(l+1)-3(10+4g_2+g_2^2)\big)\frac{\mu^2}{4m_2^2}\,,\nonumber\\
B_4 =&\ \frac{3\lambda_0}{l(l+1)}\bigg[g_2 l(l+1)\frac{\mu}{m_2} + \frac{g_2^2 -4l(l+1)}{4}\frac{\mu^2}{m_2^2}
 - \frac{g_2}{2} \frac{\mu^3}{m_2^3} + \frac{\mu^4}{4m_2^4}\bigg]\,,\nonumber\\
 B_3 =&\ 2 g_2(3-\lambda_1)\frac{\mu}{m_2} + \big( g_2^2 \lambda_2 -6(1+g_2)+ 2\lambda_1 \big)\frac{\mu^2}{m_2^2}
 + (3-\lambda_2)\bigg[2 g_2\frac{\mu^3}{m_2^3} - \frac{\mu^4}{m_2^4}\bigg]\,.
\end{align}

In order to verify the correctness of our results we will compare them with known results for limiting cases.
We restrict now to point-like particles by setting $\alpha_{Ea} = 0$.
Firstly, we perform the limit of $m_1 \rightarrow \infty $ in Eq.~(\ref{results0s12}), i.e., the limit of an infinitely heavy spinless particle. Defining
\begin{equation}
 k = (l-j)(2j+1)
\end{equation}
then Eq.~(\ref{results0s12}) leads to the nonrecoil result
\begin{align}\label{diracresult1}
	&\lim_{m_1 \rightarrow \infty} E^{(6)} = m_{2} (Z\alpha)^6 \bigg[f^{(6)}(n,j) \nonumber \\ &
	+\kappa_2\bigg(\frac{-9+19k+16k^2}{2|k|(2k-1)(2k+1)(2k+3)\,n^5}
	-\frac{3}{2k^2(2k+1)\,n^4} 
	-\frac{-3-5k+37k^2+66k^3+24k^4}{2|k|k^2(k+1)(2k-1)(2k+1)^2(2k+3)\,n^3}\bigg)\nonumber\\&
	+\kappa_2^2\bigg(\frac{3(k+1)}{2|k|(2k-1)(2k+1)(2k+3)\,n^5}-\frac{3}{2k^2(2k+1)^2\,n^4}
	-\frac{-3-5k+55k^2+120k^3+60k^4}{2|k|k^2(k+1)(2k-1)(2k+1)^3(2k+3)\,n^3}\bigg)\bigg]\, .
\end{align}
For $\kappa_2=0$ and both $l = j \pm 1/2$ the expression (\ref{diracresult1}) coincides with the Dirac result acquired by expanding the expression on the right-hand side of Eq.~(\ref{DiracRes})
up to the order $(Z\alpha)^6$. The leading recoil correction of the order
$m_{2}^{2}/m_{1}$ can be compared with the result from the literature. 
Defining the function
\begin{align}\label{f(6,1)}
&f^{(6,1)}_{1/2}(n,k) =
\frac{1}{2 n^6} - 
	\frac{-3 -2k + 14k^2 + 10 k^3}{|k| (2k-1) (2k +1) (2k +3)n^5} 
	+ \frac{3}{8 k^2 n^4} + \frac{-3 -2k + 12 k^2 + 56k^3}{8|k|k^2 (2k-1) (2k+1) (2k+3) n^3}
\end{align}
then by expanding Eq.~(\ref{results0s12}) for large $m_1$ up to the order $m_2^2/m_1$, we get the leading recoil correction
\begin{align}
	&E^{(6, 1)}=  (Z\alpha)^6 \, \frac{m_{2}^2}{m_{1}} \, \biggl[f^{(6,1)}_{1/2}(n,k) \nonumber\\&
	+\kappa_2\bigg( -\frac{-12 + 27k + 22 k^2}{|k| (2k-1) (2k+1) (2k+3)n^5} 
	+ \frac{3}{k^2 (2k+1) n^4} + \frac{-3 -5k + 49k^2 + 96k^3+36k^4}{|k|k^2 (k+1) (2k-1) (2k+1)^2(2k+3) n^3}\bigg) \nonumber\\
	&+\kappa_2^2\bigg( - \frac{9 + 11k }{2|k| (2k-1) (2k+1) (2k+3)n^5} 
	+ \frac{9}{2k^2 (2k+1)^2 n^4} + \frac{3\big(-3 -5k + 57k^2 + 128k^3+68k^4\big)}{2|k|k^2 (k+1) (2k-1) (2k+1)^3(2k+3) n^3}\bigg)\bigg]
	\, . \label{rozwinieciemoje}
\end{align}
This is in agreement with the known result from~\cite{Golosov} for $l=1$ and $\kappa_2=0$. 
For the more general case of $l>0$ the energy correction can be obtained by means of combining the expansion of Eq.~(\ref{diracrec}) up to the order $(Z\alpha)^6$  
with Eq.~(75) from~Ref.~\cite{PachuckiJentschura}. The resulting sum can be written as follows
\begin{align}
	& E^{(6, 1)}(n, j, l) = \frac{m_{2}^2}{m_{1}} (Z\alpha)^6 \biggl[ \frac{1}{2 n^6} - \frac{2}{(2j+1) n^5} 
	 + \frac{3}{2(2j+1)^2 n^4} + \frac{1}{(2j+1)^{3} n^3} \biggr] 
	+ \frac{m_{2}^{2}}{m_{1}} \frac{(Z \alpha)^2}{2 \mu^4} \left\langle \frac{\vec{L}^2}{r^4} \right\rangle \, , \label{ogolnemnadM}
\end{align}
where $\langle 1/r^4\rangle$ is given in Eq. (\ref{dodatekB:r4}). 
Comparing equation (\ref{rozwinieciemoje}) for $\kappa_2=0$ with~(\ref{ogolnemnadM}) for $l=j\pm1/2$ confirms that they agree.

The next term in the mass ratio expansion, of the order $m_{2}^{3}/m_{1}^2$ for $\kappa_2=0$, is
\begin{align}
	E^{(6, 2)}=&  (Z\alpha)^6 \, \frac{m_{2}^3}{m_{1}^2} \, \biggl[-\frac{15}{16 n^6} + 
	\frac{-9 -22k + 84k^2 + 72 k^3}{4|k| (2k-1) (2k +1) (2k +3)n^5} 
	- \frac{3(2k-1)}{8 k^2(2k+1) n^4} \nonumber \\& - \frac{3 -k -20 k^2 + 184k^3+480k^4+304k^5}{8|k|k^2 (k+1) (2k-1) (2k+1)^2 (2k+3) n^3}
	\bigg]\, , \label{recoil2}
\end{align}
which can be considered as a new result.

Secondly, we can verify that in the limiting case of Eq.~(\ref{results0s12}), when the spin-$1/2$ particle becomes infinitely heavy, we 
obtain the result in agreement with that obtained from the Klein-Gordon equation, i.e.,
\begin{equation}
	\lim_{m_2 \rightarrow \infty} E^{(6)} = m_{1} (Z\alpha)^6 f^{(6)}(n,l) \, .
\end{equation}
Similarly as in the leading relativistic correction $E^{(4)}$ this coincides with the Dirac result in Eq. (\ref{diracresult1}) by replacing $j$ by $l$ and setting $\kappa_2=0$.
Lastly, we also present the next term in the expansion of Eq.~(\ref{results0s12}) in terms of the $m_{1}^{2}/m_{2}$ ratio. Namely,
\begin{align}
	 E^{(6, 1)} =&  (Z\alpha)^6 \, \frac{m_{1}^2}{m_{2}} \, \biggl[ f^{(6,1)}_0(n,l)
	+ g_2\, \vec{L}\cdot\vec{s}_2 \biggl( -\frac{2}{l (l+1) (2 l+1) n^5} 
	 + \frac{3}{l (l+1) (2 l+1)^2 n^4} + \frac{1 + 6 l(l+1)}{l^2 (l+1)^2 (2 l+1)^3 n^3} \biggr) 	\biggr] 
	\, .\nonumber\\
\end{align}
\end{widetext}
where $f^{(6,1)}_0(n,l)$ is defined in Eq.~(\ref{f610}).

\subsection{Two particles with spins $s_1=s_2=1/2$}

Finally, we present the result for the energy $E^{(6)}$ for the case of two spin-1/2 particles.
The corresponding derivation of the first-order and second-order contributions is presented in Appendix \ref{app:spin1}.
We express the result using a set of basic spin-dependent operators. To do that we use the identities presented
in Appendix \ref{app:SO}. Moreover, to keep the formulas compact, we introduce the following symbols
\begin{align}
&\lambda_3 = -\frac{3}{(2l-1)(2l+3)}+\frac{1}{2(2l+1)^2}\,,\\
&\lambda_4 = -\frac{3}{4l(l+1)}-\frac{3}{(2l-1)(2l+3)}-\frac{3}{2(2l+1)^2}\,,\\
&\lambda_5 = \frac{3}{2l^2}-\frac{1}{2l(l+1)}+\frac{3}{2(l+1)^2}\,,\\
&\lambda_6 = \frac{3}{4l^2}-\frac{9}{4l(l+1)}+\frac{3}{4(l+1)^2}
+\frac{4}{(2l-1)(2l+3)}\nonumber\\&
-\frac{6}{(2l+1)^2}\,,\\
&\lambda_7 =-\frac{4}{(2l-1)(2l+3)}+\frac{2}{3(2l+1)^2}\,,\\
&\lambda_8 = -\frac{3}{l^2}+\frac{7}{l(l+1)}-\frac{3}{(l+1)^2}-\frac{72}{(2l-1)(2l+3)}\nonumber\\&
+\frac{20}{(2l+1)^2}\,,\\
&\lambda_9 = -\frac{1}{l^2}+\frac{3}{l(l+1)}-\frac{1}{(l+1)^2}
-\frac{12}{(2l-1)(2l+3)}\nonumber\\&
+\frac{8}{(2l+1)^2}\,,\\
&\lambda_{10} =-\frac{3}{l^2}+\frac{13}{l(l+1)}-\frac{3}{(l+1)^2}
+\frac{36}{(2l-1)(2l+3)}\nonumber\\&
+\frac{32}{(2l+1)^2}\,,\\
&\lambda_{11} = \frac{15}{l^2}-\frac{29}{l(l+1)}+\frac{15}{(l+1)^2}
-\frac{12}{(2l-1)^2}\nonumber\\&
+\frac{116}{(2l-1)(2l+3)}-\frac{96}{(2l+1)^2}
-\frac{12}{(2l+3)^2}\,,
\end{align}
in addition to $\lambda_0$, $\lambda_1$, and $\lambda_2$ defined in the previous section. 
The final result for $E^{(6)}$ of two spin-1/2 particles is then
\begin{widetext}
\begin{equation}\label{spinspin}
E^{(6)} = E^{(6)}_{s_1=s_2=0}+
\frac{\mu(Z\alpha)^6}{l(l+1)(2l-1)(2l+1)(2l+3)} \bigg[ A + B \,\vec L\cdot\vec s_1
 + C\,\vec L\cdot\vec s_2 + D\, \vec s_1\cdot\vec s_2 + F\, (L^i L^j)^{(2)} s_1^i s_2^j
\bigg]\,,
\end{equation}
where we have again subtracted the spinless result $E^{(6)}_{s_1=s_2=0}$ given by Eq.~(\ref{E6spin00}).
Using once more the definition (\ref{X}) for $X=A,B,C,D,F$ we get the following coefficients
\begin{align}
A_5 =&\
\frac{l(l+1)\,(g_1-2)}{4}\bigg(\frac{g_1+2}{2}\frac{\mu^2}{m_1^2} + (g_1-1)\frac{\mu^3}{m_1^3}\bigg)
+\frac{\mu^4}{m_1^2 m_2^2}\frac{l(l+1)}{2}
+(1\leftrightarrow2)\,,\nonumber\\
A_4 =&\ -\frac{3\lambda_0}{16} 
-\frac{3\lambda_0}{4}(g_1-2)\bigg(\frac{g_1+2}{2}\frac{\mu^2}{m_1^2}-\frac{\mu^3}{m_1^3}\bigg)
-\frac{\mu^4}{m_1^2 m_2^2}\frac{3}{8(2l+1)}\bigg(4l(l+1)
+\frac{3g_1^2 g_2^2}{16}\,
-3\bigg)+(1\leftrightarrow2)\,,\nonumber\\
A_3 =&\ \frac{\lambda_1-3}{8}
+\frac{(g_1-2)}{4}\bigg(\frac{(g_1+2)}{2}(-9+2\lambda_1)\frac{\mu^2}{m_1^2}-(-9+3g_1+2\lambda_1)\frac{\mu^3}{m_1^3}\bigg)\nonumber\\&\
+\frac{\mu^4}{m_1^2 m_2^2}\frac{1}{2}\bigg(-\frac92+\lambda_3
+\frac{g_1^2g_2^2\,\lambda_4}{16}-\lambda_4\bigg)+(1\leftrightarrow2)\,,
\end{align}
where the symbol $(1\leftrightarrow2)$ stands for the exchange of masses $m_1\leftrightarrow m_2$ and $g$-factors 
$g_1\leftrightarrow g_2$ of both particles.
For the spin-orbit contribution coefficient $B$ we get the following result, 
\begin{align}
B_5 =&\ \frac{\mu^2}{8m^2_1}\big(24(g_1-1) - 6 g_1^2 - 3 g_1 g_2(1-g_2) + 16 l(l+1)(2-g_1)\big)
+\frac{\mu^2}{8m_1 m_2} g_1 \big(48 + 3 g_2(1-g_2) - 64l(l+1)\big)\nonumber\\
&\ -
\frac{3\mu^3}{8m_1^3}\big(12(1-g_1) - g_1 g_2(1-g_2) - 16 l(l+1)(1-g_1)\big)
+\frac{\mu^4}{m_1^2 m_2^2}\bigg(-\frac92 - \frac38 g_1 g_2 + \frac{3}{32} g_1^2 g_2^2 + 6 l(l+1)\bigg)\,,\nonumber\\
B_4 =&\ \frac{3}{8l(l+1)}\bigg[\frac{\mu^2}{m^2_1}\lambda_0
\big( 2(g_1^2-1) + g_1 g_2(1-g_2) + 8l(l+1)(g_1-1)\big)
+\frac{\mu^2}{m_1 m_2} g_1 \lambda_0\big( g_2(g_2-1) + 8l(l+1)\big)
\nonumber\\
& + \frac{\mu^3}{m_1^3}  \lambda_0( 4(1-g_1) - g_1 g_2(1-g_2))
+\frac{\mu^4}{m_1^2 m_2^2}\frac{ 9 g_1^2 g_2^2 + ( g_1 g_2 + 2) \big(32 l(l+1)-24\big)}{8(2l+1)}\bigg]\,,\nonumber\\
B_3 =&\ \frac{\mu^2}{2m^2_1}\big(-6 + \big( 2g_1(g_1-1) + g_1 g_2(1-g_2) \big)\lambda_2 + (1-g_1)\lambda_5\big)
+\frac{\mu^2}{2 m_1 m_2} g_1 \big(12 - (2+g_2(1-g_2))\lambda_2-\lambda_5\big)
\nonumber\\
&\ +\frac{\mu^3}{2m_1^3}\big(12(g_1-1)
+\big(4(1-g_1)-g_1 g_2(1-g_2)\big)
\lambda_2\bigg)+\frac{\mu^4}{2m_1^2 m_2^2}\bigg(-6+\big(2+g_1 g_2\big)
\lambda_2
+\frac38 g_1^2 g_2^2\lambda_6\bigg)\,.
\end{align}
The coefficient $C$ is obtained from $B$ simply by the exchange $(1\leftrightarrow2)$. 
The spin-spin coefficient $D$
reads
\begin{align}
D_5 =&\ \frac{\mu^2}{m_1 m_2}\frac{l(l+1)\,g_1 g_2}{6}
+\frac{\mu^4}{m_1^2 m_2^2}\frac{4l(l+1)}{3}\,,\nonumber\\
D_4 =&\ -\frac{\mu^3}{m_1^2 m_2} (g_1-2)\lambda_0 - \frac{\mu^2}{2 m_1 m_2}\lambda_0 \big( 2 (1-g_1) + g_1 g_2\big)
-\frac{\mu^4}{m_1^2 m_2^2}\frac{3\,
\big(g_1^2 g_2^2-16\big)+64l(l+1)}{32(2l+1)}
+ (1\leftrightarrow2)\,,\nonumber\\
D_3 =&\ \frac{\mu^3}{m_1^2 m_2}\frac{2(g_1-2)(\lambda_1-3)}{3}
+\frac{\mu^2}{12m_1 m_2}\big(24(g_1-1) - 15 g_1 g_2 + 4\big(2(1-g_1) + g_1 g_2\big)\lambda_1\big)
\nonumber\\
&+\frac{\mu^4}{2m_1^2 m_2^2}\bigg(
\frac{1}{12}\big(g_1^2 g_2^2-16\big)\lambda_4
-6+\lambda_7\bigg)+ (1\leftrightarrow2)\,.
\end{align}

Finally, the coefficient $F$ of the tensor spin-spin interaction is the most complicated and equals
\begin{align}
F_5 =&\ \frac{1}{(2l-1)(2l+3)}\bigg[\frac{\mu^3}{m^2_1 m_2}
\frac94 (g_1-2) \big(4-6g_2+3g_1 g_2+ 16 \,l(l+1)\big)
\nonumber\\&
+\frac{\mu^2}{2m_1 m_2}\bigg(18(1-g_1) - \frac{45}{4}g_1 g_2 
+ l(l+1)\big(47 g_1 g_2 + 72 (1-g_1)\big)\bigg)\nonumber\\&
+\frac{\mu^4}{m_1^2 m_2^2}\bigg(\lambda_0(5-3g_1 g_2)(2l+1)
+6+\frac92g_1 g_2-\frac98 g_1^2g_2^2\bigg)\bigg]+(1\leftrightarrow2)\,,\\
F_4 =&\ -\frac{\mu^3}{m_1^2m_2}\frac{3(g_1-2)}{4l(l+1)(2l+1)}\big(24-18g_2 + 9 g_1 g_2 + 16l(l+1)\big)\nonumber\\&
+\frac{\mu^2}{m_1 m_2}\frac{3}{8l(l+1)(2l+1)}\big(48(g_2-1) - 15 g_1 g_2 + l(l+1)(32(g_2-1)-28 g_1 g_2)\big)\nonumber\\
&+\frac{3\mu^4}{2m_1^2 m_2^2}\bigg(\frac{48-72 g_1 g_2 + 15 g_1^2 g_2^2 + l(l+1)(12 g_1^2 g_2^2 - 64)}{16l(l+1)(2l+1)}
-\frac{3\,g_1^2g_2^2}{4\lambda_0}
\bigg)
+(1\leftrightarrow2)\,,\\
F_3 =&\ \frac{\mu^3}{4m_1^2m_2}
(g_1-2)\big(8\lambda_8
+9(g_1-2)g_2\lambda_9\big)
+\frac{\mu^2}{8m_1 m_2}\bigg(\frac{300 g_1 g_2}{\lambda_0 (2l+1)} - g_1(7g_2-4) \lambda_{10} - 8(g_1-2)\lambda_8 + 36(g_1-1) g_2\lambda_9\bigg)\nonumber\\
&+\frac{\mu^4}{2m_1^2 m_2^2}\bigg(-\lambda_{10}-\frac{12}{\lambda_0(2l+1)}
+\frac92\,g_1 g_2\bigg(\lambda_9+\frac{4}{\lambda_0(2l+1)}\bigg)
+\frac{1}{16}g_1^2g_2^2\lambda_{11}\bigg)+(1\leftrightarrow2)\,.
\end{align}

The above coefficients can be verified against the positronium limit of (\ref{spinspin}) by setting $m_1=m_2$, $Z=1$, $\kappa_1=\kappa_2=0$ and both particles being point-like.
The obtained result is in agreement with the former one from Ref.  \cite{adkinsffk}. It is worth mentioning that although we derived the result for the case of $l>1$
the final formula gives a correct result also for $l=1$. Moreover,
we would like to point out that in Ref. \cite{Zatorski} during the evaluation of Eq.~(125) for $\delta E_5$ an error was made for the case of $l>1$. Namely, in the final result 
for $\delta E^{(6)}(n, j, l, s)$  instead of the value $l$, the number $1$ was inserted in the part coming from $\delta E_5$ affecting results in Eqs. (205), (209), (213), and (217). 
After correcting this error, the results of Ref. \cite{Zatorski} are in accordance with those of Ref. \cite{Adkins2019} as well as with this work.

Exploring now the limiting case of Eq.~(\ref{spinspin}) for two point-like particles when the particle of mass $m_1$ is infinitely heavy, we set $\kappa_2=0$ and obtain the Dirac result
\begin{equation}
E^{(6,0)} = m_2 (Z\alpha)^6 \,f^{(6)}(n,j)\,,
\end{equation}
where we used the total angular momentum $\vec J = \vec L+\vec s_2$.
The first-order recoil correction given by expanding Eq.~(\ref{spinspin}) for large $m_1$ up to the order $m_2^2/m_1$, using Eq.~(\ref{f(6,1)}), reads
\begin{align}
&E^{(6, 1)}=  (Z\alpha)^6 \, \frac{m_{2}^2}{m_{1}} \, \biggl[f^{(6,1)}_{1/2}(n,k)\nonumber\\
&+\frac{g_1}{2}\bigg\{\vec s_1\cdot\vec s_2\bigg(\frac{2}{3 (2l-1)(2l+1)(2l+3)\,n^5}
-\frac{2}{l(l+1)(2l+1)^2\,n^4}
-\frac{2(-3-11l+25l^2+72l^3+36l^4)}{3l^2 (l+1)^2(2l-1)(2l+1)^3(2l+3)\,n^3}\bigg)\nonumber\\
&+\vec L\cdot\vec s_1 \bigg(-\frac{-21+32l+32l^2}{2l (l+1)(2l-1)(2l+1)(2l+3)\,n^5}
+\frac{3}{2l^2(l+1)^2\,n^4}
+\frac{-3-5l+19l^2+48l^3+24l^4}{2 l^3 (l+1)^3 (2l-1)(2l+1) (2l+3)\,n^3}\bigg)\nonumber\\
&+ (L^i L^j)^{(2)} s_1^i s_2^j \bigg(
\frac{-63+116l+116l^2}{l(l+1)(2l-1)^2 (2l+1)(2l+3)^2\,n^5}
-\frac{3(3+20l+20l^2)}{l^2 (l+1)^2(2l-1)(2l+1)^2(2l+3)\,n^4}\nonumber\\
&-\frac{-9-75l-115l^2+640l^3+2120l^4+2160 l^5+720l^6}{l^3 (l+1)^3(2l-1)^2 (2l+1)^3 (2l+3)^2\,n^3}\bigg)\bigg\}\bigg]
\end{align}
\end{widetext}
and can be considered as a new result.

\section{Summary and conclusions}

We have studied two-body systems with  constituents having arbitrary mass, arbitrary  magnetic moment, and spin $s=0$ or $ s=1/2$,
in states with angular momentum $l>1$.
The expansion of energy levels up to $\alpha^6\,\mu$ has been obtained in an analytic form and verified against special cases
where one of the particles is much heavier. The first two expansion terms in the mass ratio have a universal character. The leading term coincides
with the solution of the Klein-Gordon or Dirac equation, while the term proportional to the mass ratio does not depend on the spin of the heavier particle.
The next expansion terms are not universal and depend on both particle spins.  
In addition, the special case of positronium $m_1=m_2$, and $\kappa_1=\kappa_2=0$ was used to verify the correctness of our results. 
Although in our calculations we assumed $l>1$, our results were in agreement also for the case of $l=1$. It is not proof, but it
may indicate that the presented results are valid in the more general case of  $l>0$ for point particles.

The obtained general formulas are valid for many two-body systems, including muonic and antiprotonic atoms. This demonstrates that
precise measurements of energy levels can serve not only for the determination of fundamental constants, but also 
for the search of unknown interactions in the range not accessible in normal atoms, namely from 1 MeV  up to 100 MeV.
Although there are no definite plans to study rotational states of, for example, protonium or $\bar p \,\alpha$, the availability of a
high-precision result is crucial for planning the corresponding measurements. 

One such measurement is considered at NIST \cite{NIST}, which involves rotational states of heavy hydrogen-like ions,
with the purpose of an independent determination of the Rydberg constant.
The vacuum polarization is completely negligible, and formulas obtained in this paper will be valid also for circular states of heavy ions because
the effective expansion parameter for these states is $(Z\alpha/n)^6$. If necessary, inclusion of the $\alpha^7$ contribution for an arbitrary mass ratio 
is feasible along the lines of Ref. \cite{twlp}. This means that one may explore the potential of high accuracy results for an arbitrary two-body system
in search for physics beyond the Standard Model.

\begin{acknowledgments}
 This work was supported by the National Science Center (Poland) Grant No. 2017/27/B/ST2/02459.
 JZ would like to thank Gregory Adkins for spotting the problem with $\delta E_5$.
\end{acknowledgments}

\appendix

\section{Derivation of the effective operator $H^{(6)}$} \label{app:H6}

Using the NRQED Hamiltonian in Eq.~(\ref{ham:H2})
we derive individual first-order effective operators of the order $\alpha^6$ for the case of two spin-$1/2$ particles
with omitting all the local terms. The derivation proceeds along the same lines as in Ref.~\cite{nrqed}.
For spinless particles the effective operators are obtained by omitting the corresponding spin terms.

The first effective operator comes from the fourth term in the NRQED Hamiltonian $H_\textrm{NRQED}$ in  Eq.~(\ref{ham:H2}) and represents the
higher-order relativistic correction to the kinetic energy,
\begin{equation}
	\delta H_{0} = \frac{p_{1}^6}{16 m_{1}^5} + \frac{p_{2}^6}{16 m_{2}^5} \, .
\end{equation}

$\delta H_1$ comes from the interaction between particles when one vertex is given by the sixth and seventh terms in $H_\textrm{NRQED}$,
and the other one is $eA_0$. It leads to
\begin{align}
	\delta H_{1} =&\ \sum_{a} \frac{3}{16 m_{a}^4} \bigl(1+\frac{4\kappa_a}{3}\bigr)
	\vec{s}_{a}\cdot\{p_{a}^{2},e_{a}\vec{\cal{E}}_{a}\times \vec{p}_{a}\}
	\nonumber\\&
	+ \frac{1}{32 m_{a}^{4}}\bigg(1+\frac{s_a (s_a+1)}{3}\bigg) \,[p_{a}^2, [p_{a}^{2}, V ]] \, ,
\end{align}
where the potential $V$ is
\begin{equation}
	V = \frac{e_{1}e_{2}}{r} = - \frac{Z \alpha}{r} 
\end{equation}
and the electrostatic fields $\vec{\cal{E}}_{a}$ are defined in Eq.~(\ref{eE}).

$\delta H_2$ arises when both particles interact via the fifth term in $H_\textrm{NRQED}$.
It leads to the effective operator
\begin{align}
	\delta H_{2} =&\ \frac{e_{1}e_{2}}{4m_{1}^{2} m_{2}^{2}} (1 + 2\kappa_1)(1 + 2\kappa_2)\, 
	\nonumber\\&
	( \vec{s}_2 \times \vec{p}_{2})^i 
\frac{1}{r^3}\left( \delta^{ij} - 3 \frac{r^{i} r^{j}}{r^2} \right) ( \vec{s}_1 \times \vec{p}_{1})^j \, .
\end{align}

$\delta H_3$ arises when one of the particles interacts through part of the second term in $H_\textrm{NRQED}$, 
\begin{equation}
	\label{HFW,3A} - \frac{e}{m} \, \vec{p} \cdot \vec{A} - \frac{e}{m} \; (1+\kappa) \vec{s} \cdot \vec{B} \, ,
\end{equation}
while the other particle through part of the third term in $H_\textrm{NRQED}$,
\begin{align} -\frac{1}{8 m^3}\left( \pi^{4} - 2e\vec{s}\cdot \vec{B}\, \pi^2 - 2\pi^{2} e\vec{s}\cdot \vec{B} \right) \,.
\end{align} 
If we define static vector potentials by Eqs.~(\ref{vecA1}) and (\ref{vecA2}),
then we can write the effective operator $\delta H_3$ as 
\begin{align}	
\label{eq:operatorH3AP}
	&\delta H_3 =  \frac{1}{4} \sum_{a=1,2} \frac{1}{m_{a}^{3}} \big\{ p^{2}_a , \vec{p}_a \cdot e_{a}\vec{\cal{A}}_a 
	+ \vec{s}_a \cdot(\vec{\nabla}_a \times e_{a} \vec{\cal{A}}_a ) 
	\big\}.
\end{align}

$\delta H_4$ comes from the exchange of two transverse photons between particles. One of the particles interacts twice by the term (\ref{HFW,3A}), 
and the other one by $\frac{e^2}{2m} \, \vec{A}^2$.
The resulting effective operator can then be obtained simply by replacing the vector potential $\vec{A}$ by the static field $\vec{\cal{A}}$,
\begin{equation}
	\delta H_{4} = \sum_{a=1,2} \frac{e_{a}^2}{2 m_{a}} \, \vec{\cal{A}}_{a}^{2} \, .
\end{equation}

$\delta H_5$ corresponds to the case when one of the particles interacts through the fifth term in $H_\textrm{NRQED}$,
and the other one through (\ref{HFW,3A}) and $eA_{0}$,
\begin{equation}
\label{eq:operatorH5AP}
	\delta H_5 = \sum_{a=1,2} \, \frac{e_{a}^2\, (1+2\kappa_a)}{4 m_{a}^2} \, \vec{s}_{a} \cdot (\vec{\cal{E}}_a \times \vec{\cal{A}}_a - 
	\vec{\cal{A}}_a \times \vec{\cal{E}}_a ).
\end{equation}

$\delta H_6$ 
represents the retardation correction to the single transverse photon exchange between particles
when both particles interact through (\ref{HFW,3A}).
The corresponding $\alpha^6$ contribution is
\begin{align}
& \langle \delta H_6\rangle =  \sum_{a\neq b}\sum_b -\frac{e_a e_b}{m_a m_b} \int \frac{d^3 k}{(2\pi)^3\,2\,k^4}\bigg(\delta^{ij}-\frac{k^i k^j}{k^2}\bigg)\nonumber\\
& \bigg\langle \big(\vec p_a + \vec s_a\times\vec\nabla_a\big)^i\,e^{i\vec k\cdot\vec r_a}
(H_0-E_0)^2\big(\vec p_b + \vec s_b\times\vec\nabla_b\big)^j
\,e^{-i\vec k\cdot\vec r_b}\bigg\rangle\,.
\end{align}
Commuting out the expression $(H_0-E_0)^2$ and performing the integration we
obtain the effective operator $\delta H_6$ in Eq.~(\ref{dH6}).


$\delta H_7$ is a retardation correction in a single transverse photon exchange where one vertex is (\ref{HFW,3A}),
and the other one comes from the fifth term in $H_\textrm{NRQED}$,
\begin{equation}
- \frac{e(1+2\kappa)}{4 m^2} \; \vec{s} \cdot (\vec{E}\times \vec{p} - \vec{p}\times \vec{E}) \, .
\end{equation}
The resulting effective operator is then
\begin{align}
	&\delta H_7 = \sum_{a=1,2} \frac{e_{a}^2\,(1+2\kappa_a)}{4 m_{a}^2} \; \vec{s}_a\cdot 
	\big( \vec{\cal{E}}_a \times \vec{\cal{A}}_a - \vec{\cal{A}}_a \times \vec{\cal{E}}_a \big) 
	\nonumber\\&
	+ \frac{ie_{a}\,(1+2\kappa_a)}{8 m_{a}^3} \, \big[ \, \vec{\cal{A}}_a \cdot(\vec{p}_{a} \times \vec{s}_{a}) + (\vec{p}_{a} \times \vec{s}_{a})\cdot \vec{\cal{A}}_a, p_{a}^2 \, \big] \, .\label{eq:operatorH7AP}
\end{align}

$\delta H_8$ comes as a correction to the Coulomb interaction between particles.
One particle interacts through the eighth term in $H_\textrm{NRQED}$,
and the other one by $eA_{0}$. Then
\begin{align}
	\delta H_{8} 
		=&\ - \frac{1}{2} \sum_{a} \bigg(\alpha_{Ea}-\frac{s_a(s_a+1)}{3\,m_a^3}\bigg)  \, \frac{Z^2 \alpha^2}{r^4} \, .
\end{align}
 

Finally, the effective operator $\delta H_9$ arises when one of the particles interacts through part of the third term in $H_\textrm{NRQED}$,
\begin{equation}
 \frac{e\,\kappa}{4 m^3} \{\vec{\pi}\cdot\vec{B}, \vec{\pi}\cdot\vec{s} \}	\, ,
\end{equation}
and the second particle through (\ref{HFW,3A}), giving rise to the single transverse photon exchange without retardation. 
The resulting effective operator can be obtained simply by replacing $\vec B$ in the last equation by its static form. Namely, we can write 
\begin{equation}
\delta H_{9} = \sum_{a}
\frac{e_{a}\,\kappa_{a}}{4 m_{a}^3} \{\vec{p}_{a} \cdot \vec{\nabla}_{a}\times\vec{\cal{A}}_{a}, \vec{p}_a\cdot \vec{s}_a \} \, .
\end{equation}
This concludes the derivation of effective operators of the order $\alpha^6$ for two spin-1/2 particles for states with $l>1$. 
%

\section{Expectation values for particles with $s_1=s_2=0$} \label{app:spin0}

To simplify the expectation values 
we set $Z\alpha=1$. The final result then needs to be multiplied by a factor $(Z\alpha)^6$.
Let us denote the individual contributions  in Eq.~(\ref{E6genera}) as
\begin{equation}
E^{(6)} = \delta_1 E^{(6)} + \delta_2 E^{(6)}\,. 
\end{equation}
We start by evaluating the first-order part $\delta E^{(6)}_1$ and again stress that
we omitted all the local Dirac delta-like contributions because we are focusing on states with $l>1$.

\paragraph*{First-order contribution $\delta_1 E^{(6)}$.}
The first-order correction to the energy is the expectation value of operator $H^{(6)}$,
and for two spinless particles we use the effective operators (\ref{zestawoperatorowgeneralH6}) omitting all spins.
We separate $\delta_1 E^{(6)}$ into expectation values of individual operators appearing in $H^{(6)}$; thus, in general we have
\begin{equation}
	\delta_{1} E^{(6)} = \sum_{i=0}^{9} \delta E_{i} \, , \label{suma1raz2s00}
\end{equation}
where
\begin{equation}
	\delta E_{i} = \left\langle \delta H_{i} \right\rangle \, .
\end{equation}
However, for spinless particles the effective operators $\delta H_2$, $\delta H_5$, $\delta H_7$, and $\delta H_9$ vanish, so
the corresponding contributions to energy are zero,
\begin{equation}
	\delta E_{2}=\delta E_{5}=\delta E_{7}=\delta E_{9} = 0 
	\, .
\end{equation}

The first nonzero contribution is the higher-order relativistic correction to kinetic energy, which can be evaluated to
\begin{align}\label{dE0}
	&\delta E_{0} =
	\frac{\mu^2}{2} \biggl( \frac{1}{m_{1}^5} + \frac{1}{m_{2}^5} \biggr) \biggl(  \mu E_{0}^{3} + 
 3  \mu  E_{0}^{2} \left\langle \frac{1}{r} \right\rangle
 \nonumber \\&
 + 3  \mu  E_{0} \left\langle \frac{1}{r^2} \right\rangle 
 + \mu  \left\langle \frac{1}{r^3} \right\rangle + 
 \frac{ 1}{2}   \left\langle \frac{1}{r^4} \right\rangle \biggr) \, ,
\end{align}
where $E_0 = -\frac{\mu}{2n^2}$.
The expectation value of operators $1/r^k$ will be resolved with the help of the formulas in Appendix \ref{dodatek:pierwszyrzad}.
The next corrections are
\begin{align}\label{dE1spinzero}
	\delta E_{1} = &\ - \frac{\mu}{8}  
	\biggl( \frac{1}{m_{1}^4} + \frac{1}{m_{2}^4}\biggr) \left\langle \frac{1}{r^4} \right\rangle\,, \\
\label{dE3}
	\delta E_{3} =&\  \frac{\mu}{m_1 m_2}\biggl( \frac{1}{m_{1}^2} + \frac{1}{m_{2}^2} \biggr)\bigg\{
	2 \mu  E_{0}^{2}  \left\langle \frac{1}{r} \right\rangle 
	+
	4 \mu  E_{0}  \left\langle \frac{1}{r^2} \right\rangle\nonumber\\& \hspace{-5ex}
+  \bigl( 2 \mu 
- E_{0} l(l+1)  \bigr)
\left\langle \frac{1}{r^3} \right\rangle
+ \frac{1 }{2} \bigl( 1 
- l(l+1) \bigr)
\left\langle \frac{1}{r^4} \right\rangle\bigg\} 
	\, ,\\
\delta E_{4} = &\ \frac{ E_{0}}{m_{1} m_{2}}  \left\langle \frac{1}{r^2} \right\rangle 
+
	\frac{1}{m_{1} m_{2}} \left\langle \frac{1}{r^3} \right\rangle \nonumber\\&
+ \frac{1}{2\mu \,m_{1} m_{2}} \biggl[ 1 - \frac{3 l(l+1)}{4 }  \biggr]
\left\langle \frac{1}{r^4} \right\rangle\, . 
\end{align}
The retardation correction due to single transverse photon exchange is
\begin{widetext}
\begin{align}\label{dE6}
	&\delta E_{6} = - \frac{ E_{0}}{2 m_{1} m_{2}} \left\langle \frac{1}{r^2} \right\rangle +
	\frac{ 1}{m_{1} m_{2}} \biggl[ -\frac14 
+ \frac{\mu E_{0}(1+l(l+1))}{m_{1} m_{2}} \biggr] \left\langle \frac{1}{r^3} \right\rangle\nonumber \\
&+ \frac{1}{m_{1} m_{2}} \biggl[ -\frac{1}{8 \mu} + \frac{\mu}{m_{1} m_{2}} 
+ l(l+1) \biggl( \frac{3}{8 \mu}  + \frac{\mu}{m_{1} m_{2}} \biggr)
 \biggr]
\left\langle \frac{1}{r^4} \right\rangle 
+ \frac{3 }{2 m_{1}^2 m_{2}^2} \biggl( 1 - \frac{l^2(l+1)^2}{4} \biggr) \left\langle \frac{1}{r^5} \right\rangle \, . 
\end{align}
\end{widetext}
Finally, correction $\delta E_8$ is
\begin{equation}\label{dE8spinzero}
	\delta E_{8} = - \frac{1}{2} \, ( \alpha_{E1}+\alpha_{E2}) \, \left\langle \frac{1}{r^4} \right\rangle 
	\, ,
\end{equation}
which concludes the evaluation of the first-order expectation values for two spinless particles.

\paragraph*{Second-order contribution $\delta_2 E^{(6)}$.}
The second-order contribution of the order $(Z\alpha)^6$ is induced by the Breit Hamiltonian $H^{(4)}$, 
\begin{equation}
\delta_2 E^{(6)} = \langle H^{(4)} \frac{1}{(E_0-H_0)'}\,H^{(4)}\rangle\,.
\end{equation}
In general, such a contribution contains singularities that must be isolated by means
of regularization and then cancelled when combined with the corresponding first-order contribution.
All such singularities are, however, proportional to the Dirac delta function and thus vanish for $l>0$.
Therefore, we can ignore these singularities and write the Breit Hamiltonian for spinless particles in the form (omitting local terms)
\begin{equation}\label{breitTs}
	H^{(4)} = - A\bigg[ \frac{1}{r} \, H_0  + \, H_0 \, \frac{1}{r}\bigg] - \frac{B}{r^2}   + \frac{C}{r^3} \, ,
\end{equation}
where
\begin{align}\label{ABC}
	A =&\ \frac{\mu}{m_{1} m_{2}} + \frac{\mu^2}{2} \biggl( \frac{1}{m_{1}^3} + \frac{1}{m_{2}^3} \biggr)   \, , \\ 
	B =&\  \frac{2 \mu}{m_{1} m_{2}} + \frac{\mu^2}{2} \biggl( \frac{1}{m_{1}^3} + \frac{1}{m_{2}^3} \biggr)   \, , \\ 
	C =&\  \frac{l(l+1)}{2 m_1 m_2}  \, .\label{ABC3}
\end{align}

After elementary calculations we can express the second-order contribution in terms
of the first-order operators $\left\langle 1/r^k \right\rangle$ listed in Appendix~\ref{dodatek:pierwszyrzad}
 and~$\mathcal E_{ij}$ listed in Appendix~\ref{dodatek:drugirzad}. Then,
\begin{align}\label{Esecondspinzero}
&	\delta_{2} E^{(6)} = 4  A^2 E_{0}  \left\langle \frac{1}{r} \right\rangle^2 
	- 4  A^2 E_{0}  \left\langle \frac{1}{r^2} \right\rangle  + 
	2  A B \left\langle \frac{1}{r} \right\rangle \left\langle \frac{1}{r^2} \right\rangle \nonumber\\&
	-2  A B \left\langle \frac{1}{r^3} \right\rangle 
	- 2  A C \left\langle \frac{1}{r} \right\rangle \left\langle \frac{1}{r^3} \right\rangle 
	+ \, \biggl( 2 A C - \frac{A^2}{2\mu}  \biggr) \left\langle \frac{1}{r^4} \right\rangle\nonumber \\&
	 + 4  A^2 E_{0}^2 \,\mathcal E_{11} + 4  A B E_{0}\, \mathcal E_{12} + 
	B^2 \,\mathcal E_{22} - 4  A C E_{0} \,\mathcal E_{13} \nonumber\\ & -  2B C \,\mathcal E_{23} 
	+ \, C^2\, \mathcal E_{33} \, .
\end{align}
Combining the first-order contribution $\delta_{1} E^{(6)}$ with the second-order term $\delta_{2} E^{(6)}$,
and using the expectation values from Appendices \ref{dodatek:pierwszyrzad} and \ref{dodatek:drugirzad}
we obtain the total correction to the energy $E^{(6)}$ for two spinless particles, presented in Eq.~(\ref{E6spin00}).

\section{Expectation values for particles with $s_1=0,\,s_2=1/2$} \label{app:spin12}

The next case is when one particle is spinless and the other one has spin $s_2=1/2$. In contrast to the previous section
the effective operators $\delta H_5$ and $\delta H_7$ do not vanish, and operators $\delta H_1$, $\delta H_3$, $\delta H_4$, and $\delta H_6$, and also the Breit Hamiltonian $H^{(4)}$ now have
spin-dependent parts.

\paragraph*{First-order contribution $\delta_1 E^{(6)}$.}
The first-order contribution is given by the sum (\ref{suma1raz2s00})
with the individual contributions resolved as follows.
Contribution $\delta E_0 = \langle \delta H_0\rangle$ is the same as in the spinless case, given by Eq.~(\ref{dE0}).
The next term differs from the spinless case and is
\begin{align}
	\delta E_{1} =& \
\delta E_1^{s_1=s_2=0}- \mu  
	\biggl[\frac{1}{32 m_{2}^4} + \frac{\vec{L}\cdot\vec{s}_{2}}{m_{2}^4}\biggl( \frac{3}{4} + \kappa_2 \biggr) \biggr] \left\langle \frac{1}{r^4} \right\rangle 
	\nonumber \\&
	- \frac{ \mu  E_{0}}{m_{2}^{4}} \biggl( \frac{3}{4} + \kappa_2 \biggr) \vec{L}\cdot\vec{s}_{2} \left\langle \frac{1}{r^3} \right\rangle \, .
\end{align}
Here $\delta E_1^{s_1=s_2=0}$ stands for the spinless result (\ref{dE1spinzero}). We will use this notation throughout the following sections.

The next term $\delta E_{2}=0$ again vanishes. The remaining nonvanishing contributions are
\begin{align}
	\delta E_{3} 
=&\ \delta E_3^{s_1=s_2=0}\nonumber\\&- \frac{\mu }{m_{1} m_{2}}  E_{0} \biggl( \frac{(1+\kappa_2)}{m_{1}^2} + \frac{1}{m_{2}^2} \biggr) \vec{L}\cdot\vec{s}_{2} 
\left\langle \frac{1}{r^3} \right\rangle
\nonumber\\& 
- \frac{\mu }{m_{1} m_{2}}\biggl( \frac{(1+\kappa_2)}{m_{1}^2} + \frac{1}{m_{2}^2} \biggr) \vec{L}\cdot\vec{s}_{2}  
\left\langle \frac{1}{r^4} \right\rangle
	\, ,\\
	\delta E_{4} 
=&\ \delta E_4^{s_1=s_2=0} + \frac{1}{m_{1} m_{2}} \biggl[  \frac{1}{4 m_{2}} \, (1 + \kappa_{2})^2\nonumber\\&
- \frac{(1+\kappa_2)}{2 m_{2}} \, \vec{L}\cdot\vec{s}_{2}  \biggr]
\left\langle \frac{1}{r^4} \right\rangle 
	\, ,\\
	\delta E_{5} =&\ - \frac{(1+2\kappa_2)}{4 m_{1}m_{2}^{2}} \, \vec{L}\cdot\vec{s}_{2} \left\langle \frac{1}{r^4} \right\rangle 
	\,, \\
	\delta E_{6} 
=&\ E_6^{s_1=s_2=0} + \frac{\mu}{2m_{1} m_{2}^3} \, (1+\kappa_{2})   \, \vec{L}\cdot\vec{s}_{2}  
\left\langle \frac{1}{r^4} \right\rangle
 \, ,\\
	\delta E_{7} =&\ - \frac{(1+2\kappa_2)}{4 m_{1}m_{2}^{2}} \, \biggl( 1 + \frac{\mu}{m_{2}} \biggr) \, \vec{L}\cdot\vec{s}_{2} \left\langle \frac{1}{r^4} \right\rangle 
	\, .
\end{align}

Contribution $\delta E_8$ can be obtained from the spinless case (\ref{dE8spinzero}) by transition $\alpha_{E2}\rightarrow
\alpha_{E2}-(4m_2^3)^{-1}$,
 and $\delta E_{9} = 0$.

\paragraph*{Second-order contribution $\delta_2 E^{(6)}$.}
For the second-order contribution we repeat the derivation from the spinless case with one difference. The Breit Hamiltonian is
again expressed in the form (\ref{breitTs}) with the coefficients $A$ and $B$ being the same as in the spinless case. However, coefficient $C$ now contains
a spin-dependent part.
The derivation proceeds along the same line leading to the result (\ref{Esecondspinzero}) where for the coefficient $C$ we obtain 
\begin{equation}
	C =  \frac{\vec{L}^2}{2 m_1 m_2} + \biggl( \frac{1 + 2 \kappa_2}{2 m_{2}^2} + \frac{1 + \kappa_2}{m_1 m_2} \biggr)  \vec{L}\cdot\vec{s}_2 
\end{equation}
instead of (\ref{ABC3}). To evaluate the term $(\vec L\cdot\vec s_2)^2$ we use Eq.~(\ref{SO2}).

\section{Expectation values for particles with $s_1=s_2=1/2$}\label{app:spin1}

Finally we evaluate the most complicated case, i.e., two spin-1/2 particles, such as in antiprotonic hydrogen or positronium.
For the first-order contribution we will use the effective operators (\ref{dH0}-\ref{zestawoperatorowgeneralH6}) with both spins nonzero.

\paragraph*{First-order contribution $\delta_1 E^{(6)}$.}
The contribution to the energy of the order $(Z\alpha)^6$ for states with $l>1$ coming from the first-order operators is again expressed
in the form of series (\ref{suma1raz2s00}).
The effective operators $\delta H_i$ may contain spin-dependent terms from both particles.
The first contribution $\delta E_0$ is the same as in the spinless case, given by Eq.~(\ref{dE0}).
The next contribution is
\begin{align}
	&\delta E_{1} 
 = \delta E_1^{s_1=s_2=0}- \mu  
	\biggl[ \frac{1}{32 m_{1}^4} + \frac{1}{32 m_{2}^4} \nonumber\\&
	+ \frac{\vec{L}\cdot\vec{s}_{1}}{m_{1}^4}\biggl( \frac{3}{4} + \kappa_1 \biggr)
+\frac{\vec{L}\cdot\vec{s}_{2}}{m_{2}^4}\biggl( \frac{3}{4} + \kappa_2 \biggr) \biggr] \left\langle \frac{1}{r^4} \right\rangle \nonumber \\
&-  \mu  E_{0} \bigg[\frac{1}{m_{1}^{4}}\biggl( \frac{3}{4} + \kappa_1 \biggr) \vec{L}\cdot\vec{s}_{1}  
+\frac{1}{m_{2}^{4}}\biggl( \frac{3}{4} + \kappa_2 \biggr) \vec{L}\cdot\vec{s}_{2} \bigg]\left\langle \frac{1}{r^3}\right\rangle \, .\nonumber\\
\end{align}
Correction $\delta E_2$ does not vanish now and is
\begin{align}
&\delta E_{2} = \frac{1}{4m_1^2 m_2^2}(1+2\kappa_1)(1+2\kappa_2)\bigg[\bigg( 4\mu E_0\left\langle\frac{1}{r^3}\right\rangle\nonumber\\&
+4\mu \left\langle\frac{1}{r^4}\right\rangle -12
(l-1)(l+2) \left\langle\frac{1}{r^5}\right\rangle\bigg)\frac{(L^i L^j)^{(2)} s_1^i s_2^j}{(2l-1)(2l+3)}\nonumber\\
&+ \bigg(4\mu E_0\left\langle\frac{1}{r^3}\right\rangle
+4\mu \left\langle\frac{1}{r^4}\right\rangle -3(l-1)(l+2) \left\langle\frac{1}{r^5}\right\rangle
\bigg)\nonumber\\&
\times\frac{\vec s_1\cdot\vec s_2}{3}\bigg]\,,
\end{align}
where we have used the expectation value identity 
\begin{equation}
\bigg\langle\delta^{ij}-3\frac{r^i r^j}{r^2}\bigg\rangle 
=\frac{6}{(2l-1)(2l+3)}\big\langle (L^i L^j)^{(2)}\big\rangle\,.
\end{equation}
The next terms are
\begin{widetext}
\begin{align}
\delta E_{3} 
=&\  \delta E_3^{s_1=s_2=0} + \frac{\mu }{m_1 m_2}
\bigg[ \bigg(- E_0 \, \left\langle \frac{1}{r^3} \right\rangle
- \left\langle \frac{1}{r^4}\right\rangle \bigg)
\bigg(\bigg(\frac{1+\kappa_2}{m_1^2}+\frac{1}{m_2^2}\bigg) \vec L\cdot\vec s_2
+\bigg(\frac{1+\kappa_1}{m_2^2}+\frac{1}{m_1^2}\bigg) \vec L\cdot\vec s_1\bigg)
\nonumber\\&
+ 6\bigg(\frac{1+\kappa_2}{m_1^2}+\frac{1+\kappa_1}{m_2^2}\bigg)
\left\langle\frac{1}{r^3}\bigg(E_0 + \frac{1}{r}\bigg)\frac{( L^i L^j)^{(2)}s_1^i s_2^j}{(2l-1)(2l+3)} \right\rangle\bigg]\\
	\delta E_{4} 
 =&\ \delta E_4^{s_1=s_2=0}+ \frac{1}{m_{1} m_{2}} \biggl[ \frac{1}{4 m_{1}} \, (1 + \kappa_1)^2 
+ \frac{1}{4 m_{2}} \, (1 + \kappa_2)^2 
- \frac{(1+\kappa_1)}{2 m_{1}} \, \vec{L}\cdot\vec{s}_{1}- \frac{(1+\kappa_2)}{2 m_{2}} \, \vec{L}\cdot\vec{s}_{2}   \biggr]
\left\langle \frac{1}{r^4} \right\rangle 
	\, ,\\
	\delta E_{5} =&\ - \frac{(1+2\kappa_1)}{4 m_{1}^2m_{2}} \, \bigg[\vec{L}\cdot\vec{s}_{1}-4(1+\kappa_2)\bigg(\frac{(L^i L^j)^{(2)} s_1^i s_2^j}{(2l-1)(2l+3)}
	+\frac{\vec s_1\cdot\vec s_2}{3} \bigg) 
\bigg]\left\langle \frac{1}{r^4} \right\rangle + (1\leftrightarrow2)
	\, ,\\
	\delta E_{6} 
=&\ \delta E_6^{s_1=s_2=0}
+ \frac{\mu}{m_{1} m_{2}} \biggl[\frac{1}{2 m^2_{1}} \, (1+\kappa_{1})   \, \vec{L}\cdot\vec{s}_{1}
+ \frac{1}{2 m^2_{2}} \, (1+\kappa_{2})   \, \vec{L}\cdot\vec{s}_{2}    \biggr]
\left\langle \frac{1}{r^4} \right\rangle \nonumber \\
&
-\frac{\mu}{m_1^2 m_2^2}(1+\kappa_1)(1+\kappa_2)\bigg(\frac23 \vec s_1\cdot\vec s_2 - \frac{ (L^i L^j)^{(2)} s_1^i s_2^j}{(2l-1)(2l+3)}
\bigg)\left\langle\frac{1}{r^4}\right\rangle \, ,\\
\delta E_7 =&\ - \frac{\mu\,(1+2\kappa_1)}{4 m_1^3 m_2}\bigg[\vec L\cdot\vec s_1 + 4(1+\kappa_2)\bigg(\frac{(L^i L^j)^{(2)}s_1^i s_2^j}{(2l-1)(2l+3)} + \frac{\vec s_1\cdot\vec s_2}{3}
\bigg)\bigg]\left\langle\frac{1}{r^4}\right\rangle
+ (1\leftrightarrow2) + \delta E_5\,. 
\end{align}
Contribution $\delta E_8$ is obtained from the spinless case (\ref{dE8spinzero}) by transition $\alpha_{Ej}\rightarrow
\alpha_{Ej}-(4m_j^3)^{-1}$.
Finally,
\begin{align}
\delta E_9 =&\ \frac{\kappa_1 (1+\kappa_2)}{2m_1^3 m_2}\bigg[\bigg(8\mu E_0
 \left\langle\frac{1}{r^3}\right\rangle
+8\mu  \left\langle\frac{1}{r^4}\right\rangle
-6(l-1)(l+2)\left\langle\frac{1}{r^5}\right\rangle\bigg)\frac{( L^i L^j)^{(2)} s_1^i s_2^j}{(2l-1)(2l+3)}\nonumber\\
&+\bigg(-\frac43\mu E_0
 \left\langle\frac{1}{r^3}\right\rangle
-\frac43\mu  \left\langle\frac{1}{r^4}\right\rangle
+(l-1)(l+2)\left\langle\frac{1}{r^5}\right\rangle\bigg)\vec s_1\cdot\vec s_2\bigg]+(1\leftrightarrow2)\,.
\end{align}
The total first-order correction $\delta_{1} E^{(6)}$ is a sum $\sum_{i=0}^9 \delta E_i$.


\paragraph*{Second-order contribution $\delta_2 E^{(6)}$. } \label{second-order}
The second-order contribution in the case of two particles with spin 1/2 is more complicated than in the previous cases
due the to presence of a tensor-like term in the Breit Hamiltonian. We write the Hamiltonian $H^{(4)}$ in the form
\begin{align}\label{breitKs}
&H^{(4)} = -\frac{\mu^2}{2} \bigg(\frac{1}{m_1^3}+\frac{1}{m_2^3}\bigg) H_0^2 - \frac{\mu}{r^2}  \bigg[\frac{\mu}{2}\bigg(\frac{1}{m_1^3}+\frac{1}{m_2^3}\bigg)+\frac{2}{m_1 m_2}\bigg]
 - \mu  \bigg[\frac{\mu}{2}\bigg(\frac{1}{m_1^3}+\frac{1}{m_2^3}\bigg)+\frac{1}{m_1 m_2}\bigg] \bigg(\frac{1}{r} H_0 + H_0\frac{1}{r}\bigg)\nonumber\\&
+\frac{1}{r^3}\bigg[ \bigg(\frac{1+\kappa_1}{m_1 m_2} + \frac{1+2\kappa_1}{2m _1^2}\bigg) \vec L\cdot\vec s_1 
+ \bigg(\frac{1+\kappa_2}{m_1 m_2} + \frac{1+2\kappa_2}{2m _2^2}\bigg) \vec L\cdot\vec s_2 + \frac{\vec L^2}{2m _1 m_2}  \bigg]
+\frac{3 }{m_1 m_2} (1+\kappa_1)(1+\kappa_2) \frac{ (n^i n^j)^{(2)} s^i_1 s^j_2}{r^3}\,,\nonumber\\
\end{align}
\end{widetext}
with $(n^i n^j)^{(2)} = n^i n^j - \delta^{ij}/3$ being a symmetric, traceless tensor.
The evaluation of the second-order contribution with the Breit Hamiltonian in the form (\ref{breitKs}) proceeds along the same line as in the previous cases
using identities in Appendices \ref{dodatek:drugirzad} and \ref{app:SO} with the exception of the term
\begin{equation}\label{tensorSec}
\bigg\langle  \frac{ (n^i n^j)^{(2)} s^i_1 s^j_2}{r^3}\frac{1}{(E_0-H_0)'} \frac{ (n^k n^l)^{(2)} s^k_1 s^l_2}{r^3}\bigg\rangle\,.
\end{equation}
There are three possible contributions that correspond to three possible values of the angular momentum of the intermediate state $l'=l, l\pm2$.
Let us denote by $Q = (n^i n^j)^{(2)} s_1^i s_2^j$ and by $\hat P_l$ the projection operator into the subspace with angular momentum $l$.
Assuming that the outer states have the angular momentum $l$, we get for the diagonal term
\begin{align}
Q\,\hat P_l\,Q 
=&\ \frac{l(l+1)}{24(2l-1)(2l+3)} \bigg(1+\frac43 \vec s_1\cdot\vec s_2\bigg)
\nonumber\\&\
- \frac{1}{8(2l-1)(2l+3)} \vec L\cdot(\vec s_1+\vec s_2\big)\nonumber\\& - \frac{(2l-3)(2l+5)}{6(2l-1)^2(2l+3)^2}(L^i L^j)^{(2)} s_1^i s_2^j\,.
\end{align}
For off-diagonal terms we get
\begin{align}
Q\,\hat P_{l+2}\,Q 
=&\ \frac{(l+1) (l+2)}{16(2l+1)(2l+3)} \bigg(1+\frac43 \vec s_1\cdot\vec s_2\bigg)\nonumber\\&
+\frac{(l+2)}{8(2l+1)(2l+3)}\vec L\cdot\big(\vec s_1+\vec s_2\big)\nonumber\\
&+\frac{(l+2)}{2(2l+1)(2l+3)^2}(L^i L^j)^{(2)} s_1^i s_2^j \,,
\end{align}
and
\begin{align}
Q\,\hat P_{l-2}\,Q 
=&\ \frac{l (l-1)}{16(2l-1)(2l+1)} \bigg(1+\frac43 \vec s_1\cdot\vec s_2\bigg)\nonumber\\&
-\frac{(l-1)}{8(2l-1)(2l+1)}\vec L\cdot\big(\vec s_1+\vec s_2\big)\nonumber\\
&+\frac{(l-1)}{2(2l-1)^2(2l+1)}(L^i L^j)^{(2)} s_1^i s_2^j\,.
\end{align}

The radial part of the second-order contribution (\ref{tensorSec}) is, in the case of the diagonal term $l'=l$, equal to $\mathcal E_{33}$, Eq.~(\ref{eq:E33}).
For off-diagonal terms the Eqs. (\ref{A1}) and (\ref{A3}) are used.

\section{Expectation values of the basic first-order operators \label{dodatek:pierwszyrzad}}

Here we list the expectation values used in this work. Derivation of the expectation values for 
operators $1/r^k$ can be found in the third section of the first chapter of book \cite{BetheSalpeter}, and setting $Z\alpha=1$ they are
\begin{align}
	 & \left\langle \frac{1}{r} \right\rangle = \frac{\mu}{ n^2}, \\
	 & \left\langle \frac{1}{r^2} \right\rangle = \frac{ 2\mu^2}{\left(2l+1\right) n^3}, \\
	 & \left\langle \frac{1}{r^3} \right\rangle = \frac{2\mu^3}{l(l+1)(2l+1) n^3}, \label{wzor:r3} \\
	 & \left\langle \frac{1}{r^4} \right\rangle = 4\mu^4 \frac{3 n^2-l(1+l) }{l(l+1)(2l - 1)(2l+1)(2l+ 3) n^5}, \label{dodatekB:r4} \\
	 & \left\langle \frac{1}{r^5} \right\rangle = \frac{4 \mu^5 \big(1 - 3l(1+l) + 5 n^2\big)}{(l - 1)l(l+1)(l+2)(2l-1)(2l+1)(3+2l) n^5}.\nonumber\\
 \end{align}
These equations are valid for $l>1$.

\section{The second-order matrix elements \label{dodatek:drugirzad}}
\noindent In this section we present the expectation values of the second-order operators of the form
\begin{align}
	\mathcal{E}_{n m} = 
	 \left\langle \frac{1}{r^n} \, \frac{1}{(E_o-H_0)'} \, \frac{1}{r^m} \right\rangle\,.
\end{align}
We again set $Z\alpha=1$ for simplification of the results.
When the angular momentum of the intermediate states is the same as for the reference state then we obtain the following results:
\begin{align}
	 & \mathcal E_{11} =  - \frac{ \mu}{2 n^2}, \\
	 & \mathcal E_{12} = - \frac{2 \mu^2 }{(2l+1) n^3}, \\
	 & \mathcal E_{22} = - \mu^3  \left( \frac{6}{(2l+1)^2 \, n^4} + \frac{4}{(2l+1)^3 \, n^3} \right), \\  
	 & \mathcal E_{13} = \frac{\mu}{2l(l+1)} \left(  	2\, \mathcal E_{12} - \left\langle \frac{1}{r^2}\right\rangle \right) \, , \\
	 & \mathcal E_{23} =  \frac{\mu}{l(l+1)} \left( 	\, \mathcal E_{22} - \left\langle \frac{1}{r^3}\right\rangle \right) \, , \\
	 & \label{eq:E33} \mathcal E_{33} = \frac{\mu^2 }{l^2(l+1)^2} \, \mathcal E_{22} - \frac{\mu^2 }{l^2(l+1)^2} \left\langle \frac{1}{r^3} \right\rangle \nonumber\\&
	- \frac{3 \mu}{2 l(l+1)}  \left\langle \frac{1}{r^4}\right\rangle  \, .
\end{align}

\noindent The second-order expectation values involving the $1/r^3$ operator were derived with the help of the identity 
\begin{equation}
	\frac{1}{r^3} = \frac{\mu}{l(l+1)} \left( \frac{1}{r^2} - \left[D_r, H_0\right] \right) \, ,
\end{equation}
where $D_r = \partial_r+\frac{1}{r}$.
For off-diagonal terms with $1/r^3$ the evaluation is more complicated. In the case of $l'=l+2$ we need to calculate the expression
\begin{equation}
\bigg\langle\frac{1}{r^3}\frac{1}{(E_0 - H_{l+2})'}\frac{1}{r^3}\bigg\rangle
\end{equation}
in which we defined the radial part of the Hamiltonian as
\begin{equation}
H_l = -\frac{1}{2\mu}\bigg(\partial_r^2 + \frac{2}{r}\partial_r - \frac{l(l+1)}{r^2}\bigg)-\frac{1}{r}\,.
\end{equation}
Using the identities
\begin{align}
\frac{1}{r^3} =&\ H_l \hat\alpha - \hat\alpha H_{l+2}\,,\\
\frac{1}{r^3} =&\ \hat\beta H_l - H_{l+2}\hat\beta\,,
\end{align}
where operators $\hat\alpha$ and $\hat\beta$ are defined by
\begin{align}
\hat\alpha=&\ -\frac{\mu}{3(l+1)(l+2)} D_r -\frac{\mu(2l+3)}{3(l+1)(l+2)} \frac{1}{r}\,\nonumber\\
&+\frac{\mu^2}{3(2l+3)(l+1)(l+2)}\\
\hat \beta=&\  \frac{\mu}{3(l+1)(l+2)} D_r -\frac{\mu(2l+3)}{3(l+1)(l+2)} \frac{1}{r}\,\nonumber\\
&+\frac{\mu^2}{3(2l+3)(l+1)(l+2)}\,,
\end{align}
we can simplify the radial part to
\begin{align}\label{A1}
&\bigg\langle\frac{1}{r^3}\frac{1}{(E_0 - H_{l+2})'}\frac{1}{r^3}\bigg\rangle =  \langle\hat\alpha \,(E_0-H_{l+2})\,\hat\beta\rangle
\nonumber\\
& = -\frac{2\mu^2 (2l+3)}{9(l+1)^2(l+2)^2} E_0 \bigg\langle\frac{1}{r^2}\bigg\rangle\nonumber\\&
- \frac{\mu^2 (7+9l+3l^2)}{3(l+1)^2 (l+2)^2 (2l+3)} \bigg\langle\frac{1}{r^3}\bigg\rangle\nonumber\\&
- \frac{\mu(27 + 45l+25l^2 +4l^3)}{18(l+1)^2 (l+2)^2}\bigg\langle\frac{1}{r^4}\bigg\rangle\,
\end{align}
For the radial part with the angular momentum of the intermediate states equal to $l'=l-2$ we repeat the calculation using
\begin{align}
\frac{1}{r^3} =&\ H_l \hat\delta - \hat\delta H_{l-2}\,,\\
\frac{1}{r^3} =&\ \hat\gamma H_l - H_{l-2}\hat\gamma\,,
\end{align}
with
\begin{align}
\hat\delta=&\  -\frac{\mu}{3l(l-1)} D_r -\frac{\mu(1-2l)}{3l(l-1)} \frac{1}{r}-\frac{\mu^2}{3l(l-1)(2l-1)}\,,\nonumber\\&&\\
\hat \gamma=&\ \frac{\mu}{3l(l-1)} D_r -\frac{\mu(1-2l)}{3l(l-1)} \frac{1}{r}-\frac{\mu^2}{3l(l-1)(2l-1)}\,.\nonumber\\&&
\end{align}
Then the radial part is
\begin{align}\label{A3}
&\bigg\langle\frac{1}{r^3}\frac{1}{(E_0 - H_{l-2})'}\frac{1}{r^3}\bigg\rangle = \langle\hat\delta\,(E_0- H_{l-2})\,\hat\gamma\rangle
\nonumber\\
&= \frac{2\mu^2(2l-1)}{9(l-1)^2 l^2} E_0 \bigg\langle\frac{1}{r^2}\bigg\rangle
+\frac{\mu^2(1-3l+3l^2)}{3(l-1)^2 l^2 (2l-1)}\bigg\langle\frac{1}{r^3}\bigg\rangle\nonumber\\&
+\frac{\mu (-3+7l-13l^2 +4l^3)}{18(l-1)^2 l^2}\bigg\langle\frac{1}{r^4}\bigg\rangle\,.
\end{align}
Eqs. (\ref{A1}) and (\ref{A3}) are used for evaluation of the second-order contribution
for two spin-1/2 particles.

\section{Reduction of spin-angular operators}\label{app:SO}

To calculate the second-order contribution in $E^{(6)}$ we use the following identities to evaluate products of spin-angular operators:
\begin{align}
(\vec L\cdot\vec s_1)^2 =& \frac{l(l+1)}{4}-\frac12\vec L\cdot\vec s_1\,,\\
(\vec L\cdot\vec s_2)^2 =& \frac{l(l+1)}{4}-\frac12\vec L\cdot\vec s_2\,,\label{SO2}\\
\frac12\big\{\vec L\cdot\vec s_1,\vec L\cdot\vec s_2\big\} =& (L^i L^j)^{(2)} s_1^i s_2^j + \frac{l(l+1)}{3}\vec s_1\cdot\vec s_2\,,\nonumber\\ &\\
\frac12\big\{(L^i L^j)^{(2)} s_1^i s_2^j ,\vec L\cdot\vec s_1\big\} =& \frac{(2l-1)(2l+3)}{24} \vec L\cdot\vec s_2 \nonumber\\&- \frac34 (L^i L^j)^{(2)} s_1^i s_2^j,\\
\frac12\big\{(L^i L^j)^{(2)} s_1^i s_2^j ,\vec s_1\cdot\vec s_2\big\} =&\frac14 (L^i L^j)^{(2)} s_1^i s_2^j\,\\
(\vec s_1\cdot\vec s_2)^2 =& \frac{3}{16} - \frac12 \vec s_1\cdot\vec s_2\,.
\end{align}
The second-order contribution is then expressed as a linear combination of the  basic spin-angular operators
in the same way as the first-order contribution.

\end{document}